\documentclass[reprint,superscriptaddress,amsmath,amssymb,aps,prb,nofootinbib,floatfix]{revtex4-2}
\usepackage{amsmath}
\usepackage{graphicx}
\usepackage{bm}
\usepackage{dcolumn}
\usepackage{hyperref}

\makeatletter
\g@addto@macro\bfseries{\boldmath}
\makeatother

\newcommand{\eps}{\varepsilon}

\newcommand{\kp}{\mathbf{k}\cdot\mathbf{p}}
\newcommand{\Egap}{E_{g}}
\newcommand{\Ep}{E_{P}}
\newcommand{\Esoc}{\Delta_{\text{soc}}}
\newcommand{\Ehf}{E_{\text{HF}}}

\newcommand{\Hkp}{h_{\mathbf{k}\cdot\mathbf{p}}}

\newcommand{\Vconf}{V_{\text{ext}}}

\newcommand{\epseff}{\eps_{\text{eff}}}
\newcommand{\epsopt}{\eps_{\text{opt}}}

\newcommand{\epsin}{\eps_{\text{in}}}
\newcommand{\epsout}{\eps_{\text{out}}}

\newcommand{\Ket}[1]{ | #1 \rangle }

\newcommand{\BraKet}[2]{ \langle #1 | #2 \rangle}
\newcommand{\BraOperKet}[3]{
\langle #1 | #2 | #3 \rangle
}

\newcommand{\Vhf}{V_{\text{HF}}}

\newcommand{\Sixj}[6]{
\left\{ \begin{matrix} #1 & #2 & #3 \\ #4 & #5 & #6 \end{matrix} \right\}
}
\newcommand{\Threej}[6]{
\left( \begin{matrix} #1 & #2 & #3 \\ #4 & #5 & #6 \end{matrix} \right)
}
\newcommand{\RME}[3]{
\langle #1 \Vert \kern0.083em #2 \kern0.060em \Vert #3 \rangle
}

\begin{document}

\title{All-order correlation of single excitons in nanocrystals using a\\
 $\kp$ envelope-function approach: application to lead-halide perovskites}
 
\author{S. A. Blundell}
\email{steven.blundell@cea.fr}
\affiliation{Univ.\ Grenoble Alpes, CEA, CNRS, IRIG, SyMMES, F-38000 Grenoble,
France}

\author{C. Guet}
\email{cguet@ntu.edu.sg}
\affiliation{School of Physical and Mathematical Sciences, Nanyang Technological
University, 637371 Singapore}

\date{\today}

\begin{abstract}
We discuss a variety of many-body approaches, within effective-mass
and $\kp$ envelope-function formalisms, for calculating correlated
single excitons in semiconductor nanocrystals (NCs) to all orders
in the electron-hole Coulomb interaction. These approaches are applied
to NCs of the lead-halide perovskite CsPbBr$_{3}$, which typically
present excitons in intermediate confinement with physical observables
often strongly renormalized by correlation (e.g., radiative decay
rate enhanced by a factor of about 7 relative to a mean-field approach,
for a NC of edge length 11~nm). The many-body methods considered
include the particle-hole Bethe-Salpeter equation, configuration interaction
with single excitations, and the random-phase approximation with exchange
(RPAE), which are shown to be closely related to each other but to
treat $\kp$ corrections differently, with RPAE being the most complete
method. The methods are applied to calculate the correlation energy,
the radiative lifetime, and the long-range Coulomb contribution to
the fine structure of the ground-state exciton. In the limit of large
NC sizes, the numerical results are shown to agree well with analytical
results for this limit, where these are known. Correlated excited
states of the single exciton are used to calculate the one-photon
absorption cross section; the shape of the resulting cross-section
curve (versus laser wavelength) at threshold and up to an excitation
energy of about 1~eV is in good agreement with experimental cross
sections. The equations for the methods are explicitly adapted to
spherical symmetry (involving radial integrals and angular factors)
and in this form permit a rapid computation for systems in intermediate
confinement.
\end{abstract}

\maketitle

\section{\label{sec:introduction} Introduction}

In 2015, Protesescu \emph{et al.}~\cite{protesescu-15-psk} reported
a new class of semiconductor nanocrystal (NC) materials, all-inorganic
lead-halide perovskites CsPbX$_{3}$ (X = Cl, Br, I), having exceptional
optoelectronic properties. These NCs emit and absorb strongly, are
free from blinking, and the emission frequency can be tuned over the
whole visible range by varying the NC size and the halide composition
X (including mixtures of different halides)~\cite{protesescu-15-psk,becker-18-psk}.
This has led to numerous recent applications~\cite{chaudhary-21-psk}
to light-emitting diodes~\cite{li-16-psk,deng-16-psk}, lasers~\cite{yakunin-15-psk,pan-15-psk},
and single-photon sources~\cite{utzat-19-psk}, among others.

The typical size range of the synthesized NCs is 6--16~nm~\cite{protesescu-15-psk,chen-17-psk,brennan-17-psk,becker-18-psk},
which is comparable to or greater than the semiconductor Bohr diameter
($2a_{B}\approx6$~nm for CsPbBr$_{3}$; see Sec.~\ref{subsec:parameters}).
Consequently, excitons in these NCs are in the regime of intermediate
confinement, with partially formed bound excitons having a strong
correlation between electron and hole. This correlation can have important
consequences. For instance, the radiative decay rate of the band-edge
exciton in CsPbBr$_{3}$ is enhanced~\cite{efros-82-sqd,takagahara-87-sqd}
by factors of order 3--16 (depending on NC size in the range 6--16~nm)
compared to its value calculated assuming noninteracting carriers~\cite{becker-18-psk}
or a mean-field treatment of the carrier-carrier Coulomb interaction~\cite{nguyen-20b-psk}.
In general, special many-body techniques are necessary to treat intermediate
confinement theoretically.

In this paper, we revisit the question of correlated single excitons
in the context of inorganic lead-halide perovskite NCs. A commonly
used approach to compute exciton properties in NCs is configuration
interaction (CI)~\cite{takagahara-87-sqd,chang-98-sqd,shumway-01-sqd,tyrrell-15-sqd}.
Quantum Monte Carlo has also been used~\cite{shumway-01-sqd}. Recent
applications to NCs of CsPbX$_{3}$ have involved Hartree-Fock (HF)
and low orders of many-body perturbation theory (MBPT)~\cite{nguyen-20a-psk,nguyen-20b-psk},
and a one-parameter variational method~\cite{becker-18-psk,sercel-19-psk,tamarat-19-psk}
introduced by Takagahara~\cite{takagahara-87-sqd}.

We employ an envelope-function approach~\cite{Kira-Koch} and consider
both the effective-mass approximation (EMA) and $\kp$ models in which
the valence band (VB) and conduction band (CB) are coupled, such as
the $4\times4$ and $8\times8$ $\kp$ models~\cite{efros-00-sqd}.
The latter enable one to compute the ``$\kp$ corrections'' to the
EMA, which account for nonparabolic terms in the band dispersion as
well as VB-CB mixing induced by the finite size of the NC~\cite{ekimov-93-sqd}.
We consider approaches to correlation in which the carrier-carrier
Coulomb interaction is treated to all orders of perturbation theory.
We start from a CI expansion, a basic all-order approach, and then
generalize this to include a more complete treatment of $\kp$ corrections.
An important aspect of our formalism is that the equations are adapted
explicitly to spherical symmetry, involving 1D radial integrals and
angular factors, which leads to a computationally efficient procedure
in intermediate confinement. Applications are given to NCs of CsPbBr$_{3}$
and compared with recent experiments.

The paper is organized as follows. In Sec.~\ref{sec:formalism},
we discuss the all-order many-body formalisms that we employ. The
spherical reduction of the many-body equations to radial integrals
and angular factors is given in the Appendix. The applications to
CsPbBr$_{3}$ are then described in Sec.~\ref{sec:applications}.
First, in Sec.~\ref{subsec:parameters}, we give the material parameters
for bulk CsPbBr$_{3}$ needed as input to the EMA and $\kp$ models.
Some of these parameters remain rather uncertain at present. We also
discuss the ``quasi-cubic'' spherical confining potential used to
model the cuboid NCs of CsPbBr$_{3}$. Applications are then given
to the correlation energy (Sec.~\ref{subsec:Ecorr}), the ground-state
radiative lifetime (Sec.~\ref{subsec:lifetime}), the long-range
Coulomb contribution to the exciton fine structure (Sec.~\ref{subsec:LR-FS}),
and the one-photon absorption cross section (Sec.~\ref{subsec:1PA_xsection}).
In Secs.~\ref{subsec:Ecorr}--\ref{subsec:LR-FS} we also discuss
the analytical results that can be derived for the EMA in two cases:
noninteracting carriers and in the limit of large NC sizes. Partly
as a test of our methods, we check where possible that in the large-size
limit our numerical procedures give the expected results.

Our conclusions are given in Sec.~\ref{sec:Conclusions}. Throughout,
all formulas are given in atomic units (a.u.), $\hbar=|e|=m_{0}=4\pi\epsilon_{0}=1$.

\section{\label{sec:formalism} All-order correlated excitons}

We consider a set of interacting carriers (electrons and holes) confined
by a mesoscopic potential $\Vconf(\mathbf{r})$. The bulk band structure
is described by a $\kp$ Hamiltonian $\Hkp$ and carrier states are
expressed in terms of products of envelope and Bloch functions~\cite{Kira-Koch}.
The total Hamiltonian is
\begin{equation}
H=\sum_{ij}\{i^{\dagger}j\}\BraOperKet{i}{\Hkp+\Vconf}{j}+\frac{1}{2}\sum_{ijkl}\{i^{\dagger}j^{\dagger}lk\}\BraOperKet{ij}{g_{12}}{kl}\,,\label{eq:Hamiltonian}
\end{equation}
where $i,j,\ldots$, etc., refer to the electron states in the bands
(valence and conduction) included in the calculation, and the notation
$\{i_{1}^{\dagger}i_{2}^{\dagger}\ldots j_{1}j_{2}\ldots\}$ indicates
a normally ordered product of creation (and annihilation) operators
for electron states $i_{1},i_{2},\ldots$ (and $j,j_{2},\ldots$ ).
The Hamiltonian $\Hkp$ can can include a $\kp$ coupling term between
the VB and the CB, as in the $8\times8$ or $4\times4$ $\kp$ models~\cite{efros-00-sqd}.
Alternatively, the VB and CB can be uncoupled, as in EMA models~\cite{Kira-Koch}.
We consider both cases in the following.

In envelope-function formalisms, the Coulomb interaction is a sum
of long-range (LR) and short-range (SR) terms~\cite{Knox}, 
\begin{equation}
g_{12}=g_{12}^{\text{LR}}+g_{12}^{\text{SR}}\,.\label{eq:g12sum}
\end{equation}
The LR term has the form of the usual Coulomb interaction 
\begin{equation}
g_{12}^{\text{LR}}=\frac{1}{\varepsilon_{\text{in}}|\mathbf{r}_{1}-\mathbf{r}_{2}|}\,,\label{eq:g12LR}
\end{equation}
including a suitable dielectric constant $\varepsilon_{\text{in}}$
for the semiconductor material (which should correspond to low frequencies
and a length scale of the size of the NC). We will not do so in the
applications here, but the LR Coulomb interaction can also be generalized
using macroscopic electrostatics~\cite{Jackson} to a system of dielectrics,
where induced polarization charges form near the boundaries between
different materials---for example, in studies of the effect of the
dielectric mismatch with the environment~\cite{karpulevich-19-sqd}.

The SR term is a contact interaction~\cite{Knox}, 
\begin{equation}
g_{12}^{\text{SR}}=\gamma^{\text{SR}}\delta^{3}(\mathbf{r}_{1}-\mathbf{r}_{2})\,.\label{eq:g12SR}
\end{equation}
In discussions of exciton fine structure, the constant $\gamma^{\text{SR}}$
is proportional to the exchange Coulomb matrix element between the
Bloch states of the VB and CB \cite{Knox,[{}] [{ [Sov.\ Phys.\ JETP \textbf{33}, 108 (1971)].}] pikus-71-sqd}.
The SR term $g_{12}^{\text{SR}}$ is formally of order $(L_{\text{atom}}/L_{\text{meso}})^{2}$~\cite{Knox},
where $L_{\text{meso}}$ the mesoscopic length scale (the size of
the quantum dot or the semiconductor Bohr radius) and $L_{\text{atom}}$
is the atomistic length scale. In a finite-size system, the quantity
$L_{\text{atom}}/L_{\text{meso}}$ is also the small parameter that
appears in $\kp$ perturbation theory, $\kp\sim L_{\text{atom}}/L_{\text{meso}}$,
and the SR Coulomb interaction thus has a leading order $O[(\kp)^{2}]$.
By contrast, the LR Coulomb interaction typically gives larger contributions
of order $O(1)$ in $\kp$ perturbation theory, with the $\kp$ terms
providing small corrections (see below). An exception can occur when
the leading orders of the LR Coulomb term are suppressed for some
reason. This happens in the exciton fine structure, where the leading
LR Coulomb term is also $O[(\kp)^{2}]$ and the LR and SR terms thus
yield contributions with a similar order of magnitude (see below and
Sec.~\ref{subsec:LR-FS}). We will consider only the LR term in our
applications, $g_{12}=g_{12}^{\text{LR}}$, although the general many-body
formalism applies to the SR term as well.

We choose a single-particle basis $\Ket{i}$ for the calculations
satisfying
\begin{equation}
(\Hkp+\Vconf+U)\Ket{i}=\epsilon_{i}\Ket{i}\,,\label{eq:basis}
\end{equation}
where $U$ is a mean field, which is in principle arbitrary. Possible
choices are an independent-particle basis $U=0$, or a HF basis $U=\Vhf$,
such as a configuration-averaged HF basis~\cite{nguyen-20a-psk,nguyen-20b-psk},
which takes account of the electron-hole ($e$-$h$) interaction in
a mean-field approximation. We will take $\Vconf$ and $U$ to be
spherically symmetric. This will lead to computationally efficient
procedures in which only the radial dimension needs to be treated
numerically, while the angular dimensions can be handled analytically.
Even though NCs of inorganic perovskites are cuboid~\cite{protesescu-15-psk},
they can be well approximated for many purposes (e.g., ground-state
energies, exciton fine structure, absorption cross sections) by a
spherical ``quasi-cubic'' potential~\cite{sercel-19a-psk,nguyen-20a-psk,nguyen-20b-psk,blundell-21a-psk}.
Nonspherical terms in $\Vconf$, arising from shape corrections to
the NC or from the underlying crystal lattice, can in principle be
treated later in the calculation as perturbations.

Noninteracting or mean-field states are physically appropriate only
in the strong-confinement limit of small NC sizes $R$, where the
carrier-carrier Coulomb energy is small compared to the kinetic energy
and thus yields only a small perturbation of the independent-particle
picture. As the NC grows in size and eventually exceeds the semiconductor
Bohr radius $R>a_{B}$, partially bound excitons begin to form and
an exciton expressed in the basis (\ref{eq:basis}) acquires strong
correlation corrections~\cite{efros-82-sqd,takagahara-87-sqd}.

\begin{figure}
\includegraphics[scale=0.35]{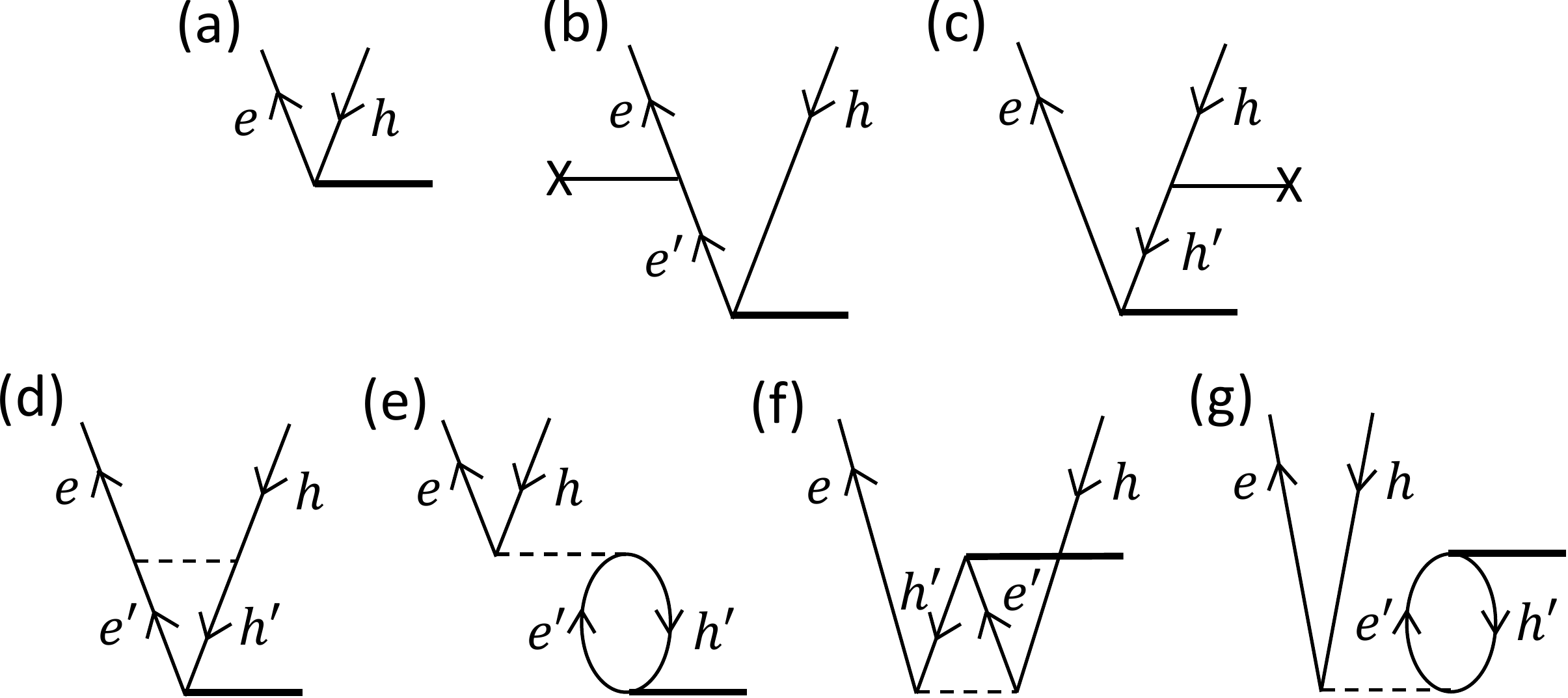}

\caption{\label{fig:Excitations}Excitations present in the various approaches
to correlated excitons. Diagrams (a)--(e) represent $\sum_{e'h'}\BraOperKet{\psi_{eh}}{H}{\psi_{e'h'}}\mathcal{X}_{e'h'}^{\alpha}$
and correspond (in order) to the five terms in Eq.~(\ref{eq:Hmatrix})
for the CIS method. The BSE approach is given by the four diagrams
(a)--(d). The RPAE method includes diagrams (a)--(e) and additionally
the diagrams (f) and (g), which represent $\sum_{e'h'}B_{eh,e'h'}\mathcal{Y}_{e'h'}^{\alpha}$
and correspond (in order) to the two terms in Eq.~(\ref{eq:Bmatrix}).
Notation: dashed horizontal line, Coulomb interaction; thick horizontal
line, all-order amplitude; horizontal line with cross, potential counter-term
$-U$.}
\end{figure}

We are interested here in studying such a confined exciton to all
orders in the Coulomb interaction $g_{12}$. As a first step, we consider
CI in the space of all single-exciton states~\cite{takagahara-87-sqd,chang-98-sqd,shumway-01-sqd,tyrrell-15-sqd}.
In this approach, a general correlated exciton state $\Ket{\alpha}$
is written 
\begin{equation}
\Ket{\alpha}=\sum_{eh}\mathcal{X}_{eh}^{\alpha}\Ket{\psi_{eh}}\,,\label{eq:CIexciton}
\end{equation}
where $\Ket{\psi_{eh}}$ is an uncorrelated single-exciton state containing
an electron $e$ and a hole $h$, 
\begin{equation}
\Ket{\psi_{eh}}=\{e^{\dagger}h\}\Ket{0}\,,\label{eq:ehexciton}
\end{equation}
with $\Ket{0}$ the effective vacuum (no carriers present in the NC).
The amplitudes $\mathcal{X}_{eh}^{\alpha}$ in Eq.~(\ref{eq:CIexciton})
are found from the CI eigenvalue problem 
\begin{equation}
\sum_{e'h'}\BraOperKet{\psi_{eh}}{H}{\psi_{e'h'}}\mathcal{X}_{e'h'}^{\alpha}=\omega_{\alpha}\mathcal{X}_{eh}^{\alpha}\,,\label{eq:CIeigenvalue}
\end{equation}
and are normalized according to 
\begin{equation}
\sum_{eh}|\mathcal{X}_{eh}^{\alpha}|^{2}=1\,.\label{eq:CInorm}
\end{equation}
To determine the matrix of the Hamiltonian $H$ in Eq.~(\ref{eq:CIeigenvalue}),
we add and subtract the mean field $U$ from Eq.~(\ref{eq:Hamiltonian}),
\begin{eqnarray}
H & = & \sum_{i}\{i^{\dagger}i\}\epsilon_{i}+\frac{1}{2}\sum_{ijkl}\{i^{\dagger}j^{\dagger}lk\}\BraOperKet{ij}{g_{12}}{kl}\nonumber \\
 &  & {}+\sum_{ij}\{i^{\dagger}j\}\BraOperKet{i}{(-U)}{j}\,,\label{eq:Hamiltonian2}
\end{eqnarray}
from which it follows that 
\begin{eqnarray}
\BraOperKet{\psi_{eh}}{H}{\psi_{e'h'}} & = & (\epsilon_{e}-\epsilon_{h})\delta_{ee'}\delta_{hh'}\nonumber \\
 &  & {}+\BraOperKet{e}{(-U)}{e'}\delta_{hh'}-\BraOperKet{h'}{(-U)}{h}\delta_{ee'}\nonumber \\
 &  & {}-\BraOperKet{eh'}{g_{12}}{e'h}+\BraOperKet{eh'}{g_{12}}{he'}\,.\label{eq:Hmatrix}
\end{eqnarray}
Many-body diagrams for $\sum_{e'h'}\BraOperKet{\psi_{eh}}{H}{\psi_{e'h'}}\mathcal{X}_{e'h'}^{\alpha}$
are shown in Figs.~\ref{fig:Excitations}(a)--(e).

To solve the eigenvalue problem (\ref{eq:CIeigenvalue}), we generate
a basis set containing all basis states $\Ket{e}$ and $\Ket{h}$
up to a high energy cutoff and compute the matrix elements $\BraOperKet{\psi_{eh}}{H}{\psi_{e'h'}}$.
It is then possible to extract the eigenvalues $\omega_{\alpha}$
and eigenvectors $\mathcal{X}_{eh}^{\alpha}$ for the correlated ground-state
exciton, $\alpha=0$, together with as many excited exciton states
$\alpha>0$ as are desired. We find that the exciton energies $\omega_{\alpha}$
are nearly independent of the choice of mean field $U=0$ or $U=\Vhf$
(there is no difference, for practical purposes, between these two
choices of $U$). The main role of a HF mean field $U=\Vhf$ in this
formalism is that fewer basis states $\Ket{e}$ and $\Ket{h}$ are
required to obtain a given precision (basis-set truncation error),
since the HF basis states already contain mean-field information about
the $e$-$h$ interaction and are more physical. We will refer to
this approach as CI singles (CIS).

The matrix element of a one-body operator $M=\sum_{ij}\{i^{\dagger}j\}\BraOperKet{i}{M}{j}$
between a correlated exciton state $\Ket{\alpha}$ and the NC ground
state $\Ket{0}$ in the CIS approach is 
\begin{equation}
\BraOperKet{\alpha}{M}{0}=\sum_{eh}(\mathcal{X}_{eh}^{\alpha})^{*}\BraOperKet{e}{M}{h}\,.\label{eq:CImxel}
\end{equation}
For example, $M$ could be the momentum operator that enters in interband
absorption and emission~\cite{Kira-Koch}.

When $\Vconf+U$ is spherically symmetric, the basis states $\Ket{e}$
and $\Ket{h}$ have exact total angular momentum quantum numbers $F_{e}$
and $F_{h}$, respectively, which couple to an exact total angular
momentum $F_{\text{tot}}$ for an exciton state~\cite{ekimov-93-sqd}.
Parity is also an exact quantum number. In the Appendix, we give the
reduction of Eqs.~(\ref{eq:CIeigenvalue})--(\ref{eq:CImxel}) to
radial integrals and angular factors for the spherical case. The computational
gain in making this reduction is not only that 1D radial integrals
are much faster to evaluate than 3D integrals, but also that the sums
over magnetic substates can be performed analytically, so that the
effective sizes of the basis set and the matrix $\BraOperKet{\psi_{eh}}{H}{\psi_{e'h'}}$
are much smaller. In the presence of nonspherical terms, these angular-momentum
quantum numbers are only approximate.

In envelope-function $\kp$ approaches, when the electron band index
changes at one vertex of a Coulomb interaction, the matrix element
is suppressed, being formally of order $O(\kp)$ in $\kp$ perturbation
theory~\cite{Kira-Koch}. Small ``$\kp$ corrections'' of this
sort correspond to nonparabolic terms in the electron dispersion relation
and to VB-CB mixing induced by the finite size of the confining potential
$\Vconf$~\cite{ekimov-93-sqd}. The $\kp$ corrections to Coulomb
matrix elements can be picked up straightforwardly in $\kp$ approaches
in which the VB and CB are coupled, such as the $8\times8$ or $4\times4$
$\kp$ models~\cite{efros-00-sqd}, where they arise as cross terms
between small and large components of the carrier wave functions in
the expression for the matrix element~\cite{nguyen-20a-psk,nguyen-20b-psk}.
Referring to Fig.~\ref{fig:Excitations}, we see that in diagram
(e), which corresponds to the last term in Eq.~(\ref{eq:Hmatrix}),
the band index necessarily changes (from the VB to the CB) at \emph{both}
vertices of the Coulomb interaction, so that this term is formally
of order $O[(\kp)^{2}]$. Diagram (e) corresponds to the $e$-$h$
exchange interaction.\footnote{Note that in atomic and molecular physics Fig.~\ref{fig:Excitations}(d)
is referred to as the ``exchange'' term and Fig.~\ref{fig:Excitations}(e)
as the ``direct'' term, which is the reverse of the convention used
in quantum-dot literature and in this paper.} In contrast, the direct $e$-$h$ Coulomb interaction in diagram
(d) has no change of band index at either vertex (for states $e$
and $e'$ in the same CB, and states $h$ and $h'$ in the same VB)
and the term is therefore formally of $O(1)$ in $\kp$ perturbation
theory. It follows that to a good approximation we can drop the last
(exchange) term $\BraOperKet{eh'}{g_{12}}{he'}$ in Eq.~(\ref{eq:Hmatrix})
and take only the first four terms. We will refer to this simplified
approach as the particle-hole Bethe-Salpeter equation or BSE approach.

\begin{figure}
\includegraphics[scale=0.37]{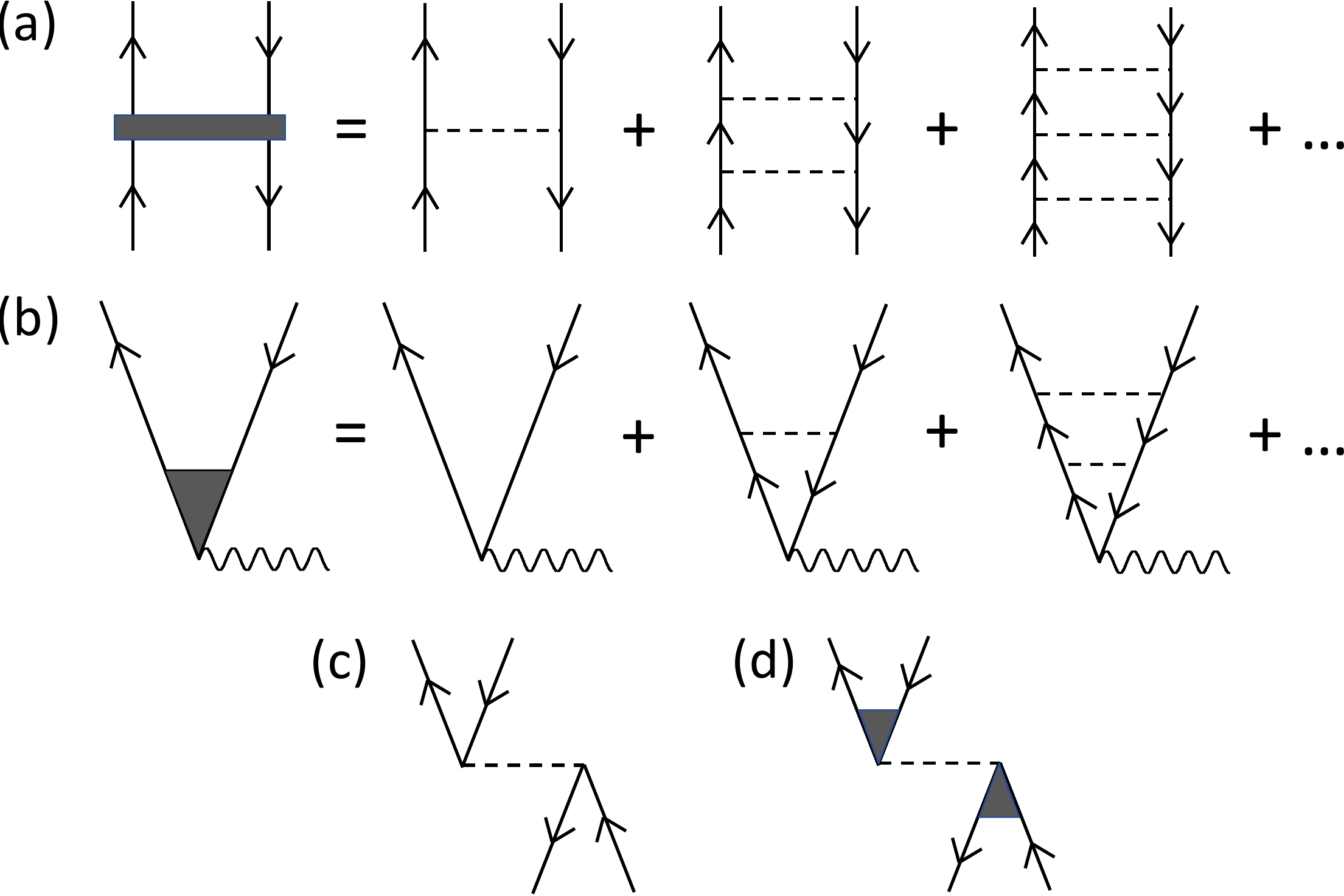}

\caption{\label{fig:ladders}(a) Effective two-body interaction formed by summing
all particle-hole ladder diagrams (arrows pointing up indicate electron
states, arrows pointing down hole states); (b) all-order vertex correction
to a one-body operator (e.g., momentum operator for interband absorption);
(c) first-order Coulomb exchange interaction for exciton fine structure;
(d) all-order Coulomb exchange interaction.}
\end{figure}

The relation to the particle-hole BSE can be seen by considering the
perturbative (iterative) solution of the CI eigenvalue equation (\ref{eq:CIeigenvalue}).
The $O(1)$ diagram shown in Fig.~\ref{fig:Excitations}(d), when
iterated, generates an effective two-body interaction given by the
sum of all particle-hole ladder diagrams~\cite{Mahan}, shown in
Fig.~\ref{fig:ladders}(a). Use of the correlated final-state wave
function $\Ket{\alpha}$ to evaluate a matrix element of a one-body
operator (\ref{eq:CImxel}) will then bring in an all-order ``vertex
correction,'' shown in Fig.~\ref{fig:ladders}(b), in which the
Coulomb ladders connect the ingoing and outgoing states of the one-body
operator. For NCs in intermediate confinement, the vertex correction
can enhance the absorption rate by large factors~\cite{efros-82-sqd,takagahara-87-sqd},
for example, by as much as 3--16 for the ground-state exciton in
NCs of the inorganic perovskite CsPbBr$_{3}$~\cite{becker-18-psk,nguyen-20b-psk}.

In the EMA, in which $\kp$ corrections are absent, the BSE approach
is entirely of order $O(1)$. In the following, we will denote this
case BSE$_{0}$. It is also possible to use the BSE approach when
the basis states are generated within a VB-CB-coupled $\kp$ approximation,
such as the $4\times4$ or $8\times8$ $\kp$ models. This approach,
which we denote BSE$_{\kp}$, brings in a subset of $\kp$ corrections
associated with the basis states. Clearly, further $\kp$ corrections
can be obtained by using the CIS approach instead, which includes
additionally the $O[(\kp)^{2}]$ exchange diagram in Fig.~\ref{fig:Excitations}(e).
However, a still more complete treatment of $\kp$ corrections is
provided by an approach analogous to the random-phase approximation
with exchange (RPAE) used in atomic and molecular physics~\cite{amusia-75-sqd}
and cluster physics~\cite{guet-92-sqd}.

In RPAE, the CI eigenvalue problem (\ref{eq:CIeigenvalue}) is replaced
by a $2\times2$ block eigenvalue problem~\cite{amusia-75-sqd} 
\begin{equation}
\left(\begin{array}{cc}
A & B\\
B^{*} & A^{*}
\end{array}\right)\left(\begin{array}{c}
\bm{\mathcal{X}}^{\alpha}\\
\bm{\mathcal{Y}}^{\alpha}
\end{array}\right)=\omega_{\alpha}\left(\begin{array}{cc}
1 & 0\\
0 & -1
\end{array}\right)\left(\begin{array}{c}
\bm{\mathcal{X}}^{\alpha}\\
\bm{\mathcal{Y}}^{\alpha}
\end{array}\right)\,.\label{eq:RPAEeigenvalue}
\end{equation}
Here, $\bm{\mathcal{X}}^{\alpha}$ and $\bm{\mathcal{Y}}^{\alpha}$
are vectors of amplitudes $\mathcal{X}_{eh}^{\alpha}$ and $\mathcal{Y}_{eh}^{\alpha}$
in the space of uncorrelated excitons $(e,h)$, and the matrix $A$
is identical to that appearing in the CIS eigenvalue problem (\ref{eq:CIeigenvalue}),
\begin{equation}
A_{eh,e'h'}=\BraOperKet{\psi_{eh}}{H}{\psi_{e'h'}}\,,\label{eq:Amatrix}
\end{equation}
which is given in detail by Eq.~(\ref{eq:Hmatrix}). The matrix $B$
is 
\begin{equation}
B_{eh,e'h'}=-\BraOperKet{ee'}{g_{12}}{h'h}+\BraOperKet{ee'}{g_{12}}{hh'}\,.\label{eq:Bmatrix}
\end{equation}
RPAE thus includes the dominant correlation diagrams of the BSE approach,
Figs.~\ref{fig:Excitations}(a)--(d), the additional exchange diagram
of CIS, Fig.~\ref{fig:Excitations}(e), and two further diagrams
associated with the $B$ matrix, shown in Figs.~\ref{fig:Excitations}(f)
and (g). These two diagrams are both of order $O[(\kp)^{2}]$ in $\kp$
perturbation theory. Physically, the $B$ matrix accounts for two-particle/two-hole
correlations in the ground state~\cite{amusia-75-sqd}. The RPAE
eigenvector is normalized according to 
\begin{equation}
\sum_{eh}(|\mathcal{X}_{eh}^{\alpha}|^{2}-|\mathcal{Y}_{eh}^{\alpha}|^{2})=1\,,\label{eq:RPAEnorm}
\end{equation}
and the matrix element (\ref{eq:CImxel}) in RPAE becomes 
\begin{equation}
\BraOperKet{\alpha}{M}{0}=\sum_{eh}[(\mathcal{X}_{eh}^{\alpha})^{*}\BraOperKet{e}{M}{h}+(\mathcal{Y}_{eh}^{\alpha})^{*}\BraOperKet{h}{M}{e}]\,.\label{eq:RPAEmxel}
\end{equation}
The angular reduction of Eqs.~(\ref{eq:RPAEeigenvalue})--(\ref{eq:RPAEmxel})
for a spherically symmetric potential is given in the Appendix.

Note that the many-body formalisms BSE, CIS, and RPAE, presented in
Eqs.~(\ref{eq:CIexciton})--(\ref{eq:RPAEmxel}), apply to any single-particle
basis $\Ket{i}$ and not just to the envelope-function basis that
we use in our applications. In Eq.~(\ref{eq:basis}), the term $\Hkp+\Vconf$
can be reinterpreted as any suitable effective single-particle Hamiltonian
describing states of the finite-size NC.

\section{\label{sec:applications}Application to perovskite nanocrystals}

\subsection{\label{subsec:parameters}Parameters and model}

Inorganic lead-halide perovskites are ``inverted'' direct-gap semiconductors
having a $p_{1/2}$-like CB and an $s$-like VB. The VB maximum ($R_{6}^{+}$)
and CB minimum ($R_{6}^{-}$) lie at the $R$ point of the Brillouin
zone~\cite{protesescu-15-psk,becker-18-psk}. This VB-CB pair can
be described by the $4\times4$ $\kp$ model~\cite{even-14b-psk,yang-17-psk,becker-18-psk}
or used for EMA calculations. The $p_{1/2}$-like CB is split by spin-orbit
coupling from a higher-lying $p_{3/2}$-like band, whose minimum ($R_{8}^{-}$)
lies about 1~eV above the minimum of the $p_{1/2}$-like band~\cite{becker-18-psk}.
The $s$-like VB together with the $p_{1/2}$- and $p_{3/2}$-like
CBs can be described by an extended $8\times8$ $\kp$ model~\cite{efros-00-sqd}.

The bulk CsPbBr$_{3}$ material parameters that we use are summarized
in Table~\ref{tab:parameters}. The bandgap $\Egap$ and reduced
effective mass $\mu^{*}=m_{e}^{*}m_{h}^{*}/(m_{e}^{*}+m_{h}^{*})$
of the $s$-like VB and $p_{1/2}$-like CB were measured by Yang \emph{et
al}.~\cite{yang-17-psk}\ for the orthorhombic phase of CsPbBr$_{3}$
at cryogenic temperatures. While $\mu^{*}$ is known experimentally,
the individual effective masses $m_{e}^{*}$ and $m_{h}^{*}$ are
not. However, there is theoretical~\cite{becker-18-psk,protesescu-15-psk}
and experimental~\cite{fu-17a-psk} evidence that $m_{e}^{*}$ and
$m_{h}^{*}$ are approximately equal in inorganic lead-halide perovskites,
so we will assume $m_{e}^{*}=m_{h}^{*}$. The spin-orbit coupling
parameter $\Esoc$ is defined as the energy splitting of the $p_{3/2}$-
and $p_{1/2}$-like CBs at the $R$ point; we estimate $\Esoc$ from
a fit to the experimental absorption spectra (see Sec.~\ref{subsec:1PA_xsection}).
Finally, the ``effective'' dielectric constant $\epseff$ was inferred
by Yang \emph{et al.}~\cite{yang-17-psk} from the bulk exciton binding
energy (see also Ref.~\cite{shcherbakov-wu-21-psk}). Since $\epseff$
applies to a length scale of order the semiconductor Bohr radius $a_{B}$,
we use $\epsin=\epseff$ to screen the carrier-carrier Coulomb interactions
{[}Eq.~(\ref{eq:g12LR}){]}.

The Kane parameter $\Ep$ has also not been measured. Estimates of
$\Ep$ were made in Ref.~\cite{nguyen-20a-psk} based on the $4\times4$
and $8\times8$ $\kp$ models discussed above, together with the assumption
that remote bands (those not included in the model) make zero contribution.
The resulting values $\Ep^{(4\times4)}$ and $\Ep^{(8\times8)}$ are
given in Table~\ref{tab:parameters} and can be seen to differ significantly.
Given that the only difference between the two values is the inclusion
of the $p_{3/2}$-like CB in $\Ep^{(8\times8)}$, it seems likely
that other remote bands may make further significant contributions
to $\Ep$. An estimate using density-functional theory (DFT) \cite{becker-18-psk}
found $\Ep=39.9$~eV, which seems quite high. As in Ref.~\cite{nguyen-20a-psk},
we take the view that $\Ep$ is presently uncertain, a conservative
range being $10\,\text{eV}\leq\Ep\leq40\,\text{eV}$. We will assume
a value $\Ep=20$~eV in our applications that is intermediate between
$\Ep^{(4\times4)}$ and $\Ep^{(8\times8)}$. 

\begin{table}[tb]
\caption{\label{tab:parameters}Bulk material parameters for CsPbBr$_{3}$
used in this paper. $\Ep^{(4\times4)}$ and $\Ep^{(8\times8)}$ are
estimates of the Kane parameter derived in Ref.~\protect\cite{nguyen-20a-psk}
from the $4\times4$ and $8\times8$ $\mathbf{k}\cdot\mathbf{p}$
models. The parameters $\epseff$ and $\epsopt$ are the effective
and optical dielectric constants of CsPbBr$_{3}$, respectively, and
$\epsout$ is the optical dielectric constant of the surrounding medium.
$\Esoc$ is the spin-orbit coupling parameter of the $8\times8$ $\kp$
model. Further explanation is given in Sec.~\ref{subsec:parameters}.}

\begin{ruledtabular}
\begin{tabular}{ld}
 & \multicolumn{1}{c}{CsPbBr$_{3}$}\\
\hline 
$\mu^{*}$ ($\times m_{0}$) & 0.126\footnotemark[1]\\
$m_{e}^{*}=m_{h}^{*}$ ($\times m_{0}$) & 0.252\\
$\Egap$ (eV) & 2.342\footnotemark[1]\\
$\Ep^{(4\times4)}$ (eV) & 27.9\footnotemark[2]\\
$\Ep^{(8\times8)}$ (eV) & 16.4\footnotemark[2]\\
$\Ep$ (eV) & 20.0\\
$\Esoc$ (eV) & 0.8\footnotemark[3]\\
$\epseff$ & 7.3\footnotemark[1]\\
$\epsopt$ & 4.84\footnotemark[4]\\
$\varepsilon_{\text{out}}$ & 2.4\footnotemark[5]\\
\end{tabular}

\footnotetext[1]{Reference~\cite{yang-17-psk}}

\footnotetext[2]{Reference~\cite{nguyen-20a-psk}}

\footnotetext[3]{From a fit to experimental absorption spectra (see
Sec.~\ref{subsec:1PA_xsection}).}

\footnotetext[4]{Reference~\cite{dirin-16-psk}, for a wavelength
of 500~nm.}

\footnotetext[5]{Applies to toluene.}
\end{ruledtabular}

\end{table}

NCs of CsPbBr$_{3}$ are cuboid~\cite{protesescu-15-psk}. For the
above material parameters, one finds $2a_{B}=6.1$~nm, implying that
NCs with edge lengths $L$ in the experimentally interesting range
$L=6$--16~nm are in the regime of intermediate confinement, with
strongly correlated excitons.

We take the confining potential to be a spherical well with infinite
walls 
\begin{equation}
\Vconf^{\text{sph}}(r)=\left\{ \begin{matrix}0\text{, if }r<R\\
\infty\text{, otherwise}
\end{matrix}\right.\,.\label{eq:sphericalWell}
\end{equation}
The radius $R$ is chosen to be 
\begin{equation}
R=L/\sqrt{3}\,,\label{eq:radiusL}
\end{equation}
which ensures that the noninteracting ground-state energy of a carrier
(hole or electron) matches that in a cubic well with edge length $L$~\cite{sercel-19a-psk,nguyen-20a-psk}.
This quasi-cubic spherical potential can be shown to reproduce well
other properties of a cubic well, such as the Coulomb energy of the
$1S_{e}$-$1S_{h}$ exciton~\cite{nguyen-20a-psk}, the correlation
energies of the trion and biexciton~\cite{nguyen-20a-psk}, and the
one-~\cite{nguyen-20b-psk} and two-photon~\cite{blundell-21a-psk}
absorption cross sections.

\subsection{\label{subsec:Ecorr}Correlation energy of ground-state exciton}

An example of a calculation of the correlation energy is given in
Table~\ref{tab:Ecorr}. We here use the BSE$_{0}$ method (in the
EMA) with a configuration-averaged HF basis set~\cite{nguyen-20a-psk,nguyen-20c-psk}
$U=\Vhf$. The states of the basis set are cut off at principal quantum
numbers $n_{\text{max}}=12$ in each angular-momentum channel ($s$,
$p_{1/2}$, $p_{3/2}$, \ldots , etc.). We then perform a series of
calculations with the orbital angular momentum of the basis set truncated
at $l_{\text{max}}=K$ for all $K$ in the range $1\leq K\leq12$,
giving a set of total exciton energies $E_{\text{tot}}(K)$. For $K\geq2$,
the partial-wave contribution $\delta E(K)$ for orbital angular momentum
$K$ is defined as $\delta E(K)=E_{\text{tot}}(K)-E_{\text{tot}}(K-1)$.
Since we define the correlation energy as the difference between the
total energy and the configuration-averaged HF energy, $E_{\text{corr}}=E_{\text{tot}}-\Ehf$,
we can take the contribution for $K\leq1$ to be $\delta E(K\leq1)=E_{\text{tot}}(1)-\Ehf$.

\begin{table}
\caption{\label{tab:Ecorr}Contribution $\delta E(K)$ of individual partial
waves to the correlation energy of the ground-state $1S_{e}$-$1S_{h}$
exciton ($F_{\text{tot}}=1$), for NCs of CsPbBr$_{3}$ with edge
lengths $L=9$~nm and 12~nm. Calculations are performed with the
BSE$_{0}$ method using an EMA basis set. The correlation energy is
here defined as the difference between the total exciton energy and
the exciton energy in the configuration-averaged HF approximation
(see Ref.~\cite{nguyen-20a-psk} for more details), $E_{\text{corr}}=E_{\text{tot}}-E_{\text{HF}}$.
We find $E_{\text{HF}}=0.08756199$~Ha for $L=9$~nm and $E_{\text{HF}}=0.08640365$~Ha
for $L=12$~nm. Notation: $K$ is the partial wave (see text); `extrap.'\ is
the extrapolated sum of terms for $K=13$ to infinity. Units: mHa.}

\begin{ruledtabular}
\begin{tabular}{ldd}
$K$ & \multicolumn{1}{c}{$L=9$~nm} & \multicolumn{1}{c}{$L=12$~nm}\\
\hline 
$\leq1$ & -0.02624 & -0.02782\\
2 & -0.21087 & -0.21166\\
3 & -0.06176 & -0.07328\\
4 & -0.02343 & -0.03096\\
5 & -0.01053 & -0.01491\\
6 & -0.00534 & -0.00792\\
7 & -0.00297 & -0.00455\\
8 & -0.00177 & -0.00277\\
9 & -0.00112 & -0.00178\\
10 & -0.00074 & -0.00119\\
11 & -0.00050 & -0.00082\\
12 & -0.00036 & -0.00058\\
extrap. & -0.00123 & -0.00206\\
Total & -0.34683(1) & -0.38030(1)\\
\end{tabular}
\end{ruledtabular}

\end{table}

From Table~\ref{tab:Ecorr}, one sees that the correlation energy
of the ground-state exciton $1S_{e}$-$1S_{h}$ is dominated by the
contribution $K=2$, but higher partial waves also make significant
contributions. Asymptotically for large $K$, one finds $\delta E(K)\sim1/K^{3}$,
which enables us to make an estimate of the contribution from $K=13$
to infinity. The estimated numerical error from the principal-quantum-number
cutoff and partial-wave extrapolation is given in the final line.
Note that, at the level of approximation used here (EMA and BSE$_{0}$),
the two possible total angular momenta $F_{\text{tot}}=0$ and 1 of
the ground-state $1S_{e}$-$1S_{h}$ exciton should be degenerate.
However, it turns out that the partial-wave contributions for $F_{\text{tot}}=0$
and 1 are slightly different (e.g., for an edge length $L=9$~nm,
$\delta E(2)=-0.21087$~mHa for $F_{\text{tot}}=1$ and $\delta E(2)=-0.17266$~mHa
for $F_{\text{tot}}=0$). Nevertheless, we find that the final extrapolated
energy for the two different values of $F_{\text{tot}}$ agree to
within the estimated error shown in the table (which applies to $F_{\text{tot}}=1$),
thus providing a useful test of the numerics.

A feature of this approach is that the partial-wave expansion becomes
more slowly convergent as the NC size $L$ increases. This can be
seen in Table~\ref{tab:Ecorr}, where the relative contribution to
the correlation energy from the extrapolated terms $K\geq13$ is greater
for $L=12$~nm than for $L=9$~nm. In fact, the method eventually
becomes unwieldy for edge lengths $L\agt50$~nm owing to the slow
convergence, but for intermediate confinement in the experimentally
interesting size range $6\,\text{nm}\leq L\leq16\,\text{nm}$, one
can readily achieve energies converged to a fractional error of 10$^{-3}$
or better in a few seconds of calculation on a single core.

\begin{figure}
\includegraphics[scale=0.5]{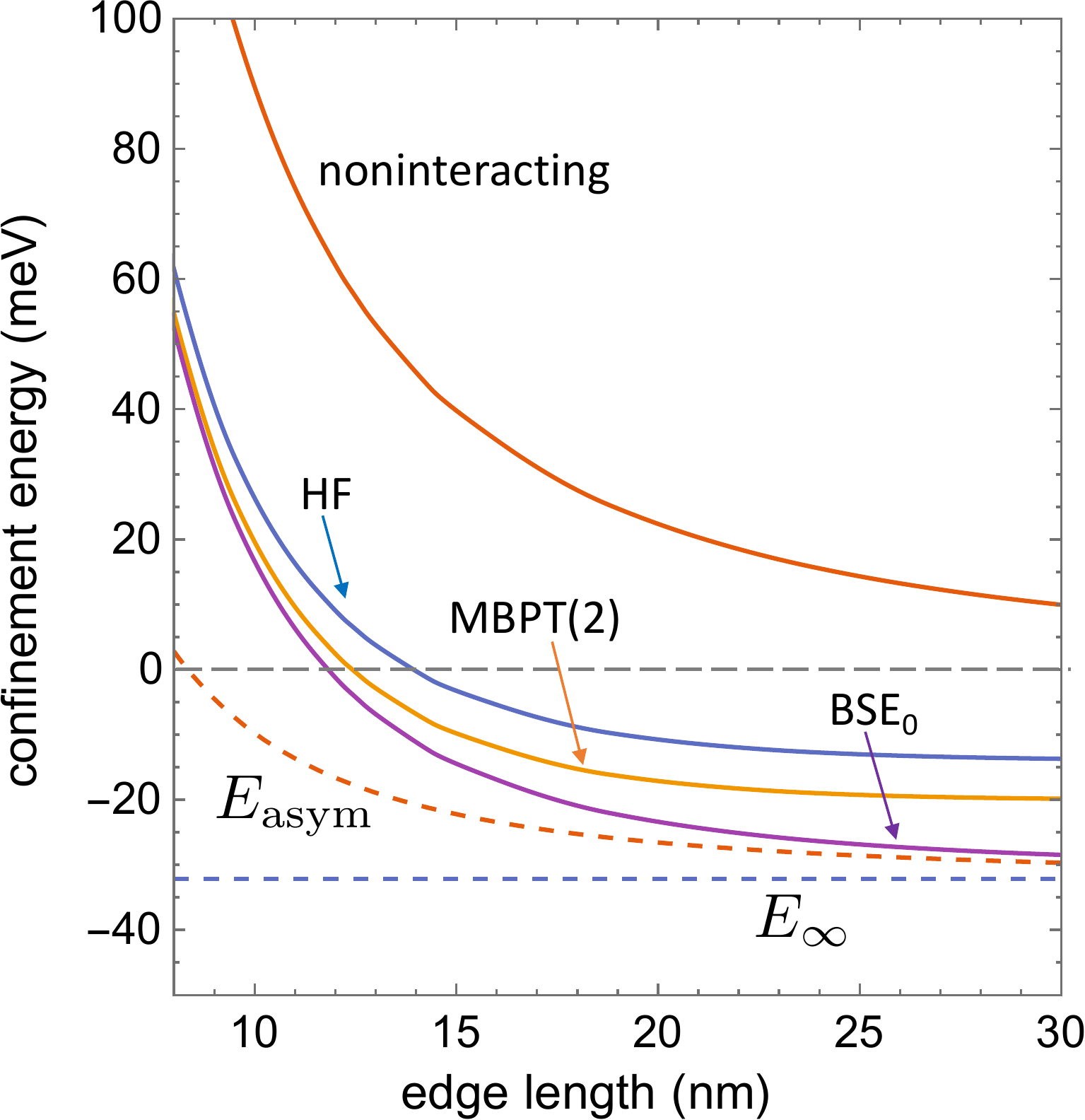}

\caption{\label{fig:conf_energy}Ground-state exciton confinement energy {[}Eq.~(\ref{eq:Econf}){]}
for a NC of CsPbBr$_{3}$ vs.\ NC edge length, in various approximations.
All approaches assume the EMA. Continuous lines (top to bottom): `noninteracting',
noninteracting carriers; HF, configuration-averaged Hartree-Fock (following
Ref.~\protect\cite{nguyen-20a-psk}); MBPT(2), many-body perturbation
theory up to second order (following Ref.~\protect\cite{nguyen-20a-psk});
BSE$_{0}$, particle-hole Bethe-Salpeter equation. Dashed lines (top
to bottom): $E_{\text{asym}}$, the large-$R$ asymptotic form of
the confinement energy, Eq.~(\ref{eq:Easym}); $E_{\infty}$, the
$R\rightarrow\infty$ limit of the confinement energy, Eq.~(\ref{eq:Einfinity}).}
\end{figure}

Figure~\ref{fig:conf_energy} shows the ground-state confinement
energy in various many-body approximations (but all using the EMA),
as a function of the NC edge length for $L=8$--30~nm. The confinement
energy here is defined as 
\begin{equation}
E_{\text{conf}}=E_{\text{tot}}-\Egap\,.\label{eq:Econf}
\end{equation}
Two cases can be handled analytically in the EMA. First, for noninteracting
carriers, the ground-state single-particle wave function is 
\begin{equation}
\psi_{1S}(\mathbf{r})=(2\pi R)^{-1/2}\frac{1}{r}\sin\left(\frac{\pi r}{R}\right)\,,\label{eq:gs_wave_function}
\end{equation}
and the noninteracting confinement energy of a $1S_{e}$-$1S_{h}$
exciton is 
\begin{equation}
E_{\text{non}}=\frac{\pi^{2}}{2\mu R^{2}}\,,\label{eq:e_non}
\end{equation}
which tends to zero in the bulk limit $R\rightarrow\infty$.

The bulk limit for interacting particles can also be handled analytically
in the EMA, by solving a two-body Schr\"odinger equation~\cite{Mahan},
\begin{align}
 & \left(-\frac{1}{2m_{e}^{*}}\nabla_{e}^{2}-\frac{1}{2m_{h}^{*}}\nabla_{h}^{2}+\frac{1}{\epsin|\mathbf{r}_{e}-\mathbf{r}_{h}|}\right)\Psi(\mathbf{r}_{e},\mathbf{r}_{h})\nonumber \\
 & \quad\quad\quad\quad\quad\quad\quad\quad\quad=\left(E_{\text{tot}}-\Egap\right)\Psi(\mathbf{r}_{e},\mathbf{r}_{h})\,,\label{eq:2body_eqn}
\end{align}
where $\Psi(\mathbf{r}_{e},\mathbf{r}_{h})$ goes to zero at the NC
boundary. In Eq.~(\ref{eq:2body_eqn}), the electron and hole are
effectively regarded as different species of particle, with one particle
of each type present. It follows that a CI solution of Eq.~(\ref{eq:2body_eqn})
is formally identical to the method we called BSE$_{0}$ in Sec.~\ref{sec:formalism}.
In particular, since the electron and hole are regarded as distinguishable,
the wave function $\Psi(\mathbf{r}_{e},\mathbf{r}_{h})$ is not required
to be antisymmetric under interchange of electron and hole coordinates,
and so the final exchange term in Eq.~(\ref{eq:Hmatrix}) is absent,
as in the definition of the BSE$_{0}$ method. Also, as mentioned
above, the spin-independence of the Hamiltonian has the consequence
that the solutions for $F_{\text{tot}}=0$ and 1 are degenerate.

In the large-$R$ limit, Eq.~(\ref{eq:2body_eqn}) can be solved
using relative $\mathbf{r}=\mathbf{r}_{e}-\mathbf{r}_{h}$ and center-of-mass
(CM) $\mathbf{R}_{\text{CM}}=(m_{e}^{*}\mathbf{r}_{e}+m_{h}^{*}\mathbf{r}_{h})/(m_{e}^{*}+m_{h}^{*})$
coordinates, 
\begin{equation}
\Psi(\mathbf{r}_{e},\mathbf{r}_{h})=\psi_{\text{CM}}(\mathbf{R}_{\text{CM}})\psi_{\text{rel}}(\mathbf{r})\,.\label{eq:2body_soln}
\end{equation}
This solution represents a bound exciton with CM fluctuations. We
are interested here in the ground state. The CM wave function $\psi_{\text{CM}}(\mathbf{R}_{\text{CM}})$
then has the same form as Eq.~(\ref{eq:gs_wave_function}), and the
relative wave function $\psi_{\text{rel}}(\mathbf{r})$ is the $1s$
hydrogen-like solution for effective mass $\mu$ (and a dielectric
constant $\epsin$). The large-$R$ asymptotic confinement energy,
including the CM contribution, is 
\begin{equation}
E_{\text{asym}}(R)=\frac{\pi^{2}}{2(m_{e}^{*}+m_{h}^{*})R^{2}}-\frac{\mu}{2\epsin^{2}}\,.\label{eq:Easym}
\end{equation}
In the limit $R\rightarrow\infty$, the confinement energy is the
bulk $1s$ exciton binding energy, 
\begin{equation}
E_{\infty}=-\frac{\mu}{2\epsin^{2}}\,,\label{eq:Einfinity}
\end{equation}
which is $E_{\infty}=-32.17$~meV for the parameters in Table~\ref{tab:parameters}.

Figure~\ref{fig:conf_energy} shows that configuration-averaged HF
picks up only about 45\% of the exciton binding energy in the large-$R$
limit. This is improved to about 65\% using MBPT up to 2nd order.
At $L=30$~nm, however, the all-order BSE$_{0}$ solution is within
only 4\% of its asymptotic value $E_{\text{asym}}(R)$~(\ref{eq:Easym});
at this size, the CM energy in Eq.~(\ref{eq:Easym}) is still significant.
We also see that 2nd-order MBPT gives a good account of the correlation
energy for smaller NC sizes $L\alt10$~nm, including the strong-confinement
limit. For larger sizes $10\text{ nm}\alt L\alt16\text{ nm}$ of experimental
interest, however, BSE$_{0}$ is significantly different from HF,
MBPT, and $E_{\text{asym}}(R)$; all-order approaches thus become
the preferred method of calculation for these sizes.

A discussion of $\kp$ corrections to the correlation energy will
be left to Sec.~\ref{subsec:LR-FS}. These corrections in general
lift the degeneracy between the solutions for $F_{\text{tot}}=0$
or 1 and contribute to the exciton fine structure.

\subsection{\label{subsec:lifetime}Radiative lifetime of the ground-state exciton}

The radiative lifetime $\tau_{\alpha}$ of an exciton state $\alpha$
is given by~\cite{efros-82-sqd,nguyen-20b-psk} 
\begin{equation}
\frac{1}{\tau_{\alpha}}=\frac{4}{9}\frac{n_{\text{out}}\omega_{\alpha}}{c^{3}}f_{\varepsilon}^{2}\left|M_{\alpha}\right|^{2}\,,\label{eq:tau_alpha}
\end{equation}
where $n_{\text{out}}=\sqrt{\epsout}$ is the refractive index of
the surrounding medium, with $\epsout$ its optical dielectric constant,
$\omega_{\alpha}$ is the frequency of the emitted photon (total exciton
energy), and $M_{\alpha}=\RME{\alpha(F_{\text{tot}})}{M}{0}$ is the
reduced momentum matrix element, which is given by Eq.~(\ref{eq:RPAEredmxel}).
A selection rule for one-photon emission requires $F_{\text{tot}}=1$,
otherwise $M_{\alpha}=0$. The quantity $f_{\varepsilon}$ is the
dielectric screening factor, defined as the ratio of photon electric
field inside the NC to that at infinity, which we take to have the
value for a sphere~\cite{Jackson} (see Refs.~\cite{becker-18-psk}
and \cite{nguyen-20b-psk} for further discussion), 
\begin{equation}
f_{\varepsilon}^{\text{sph}}=\frac{3\varepsilon_{\text{out}}}{\varepsilon_{\text{opt}}+2\varepsilon_{\text{out}}}\,.\label{eq:feps-sphere}
\end{equation}

As in the previous section, two cases of interest can be handled analytically
within the EMA. First, for noninteracting carriers the reduced momentum
matrix element for the ground-state $1S_{e}$-$1S_{h}$ exciton is
given by~\cite{nguyen-20b-psk} 
\begin{equation}
\left|M_{\alpha}\right|^{2}=\left|\BraKet{1S_{e}}{1S_{h}}\right|^{2}\left|\RME{J_{\text{CB}}}{p}{J_{\text{VB}}}\right|^{2}=\Ep\,.\label{eq:redmxel_non}
\end{equation}
Here, $\BraKet{1S_{e}}{1S_{h}}=1$ is the overlap of envelope functions,
and $\RME{J_{\text{CB}}}{p}{J_{\text{VB}}}$ is the reduced interband
momentum matrix element between Bloch states; this satisfies $\left|\RME{J_{\text{CB}}}{p}{J_{\text{VB}}}\right|^{2}=\Ep$,
which can be regarded as the definition of the Kane parameter $\Ep$~\cite{nguyen-20b-psk}.
Hence, for noninteracting particles the lifetime of the $1S_{e}$-$1S_{h}$
exciton is 
\begin{equation}
\frac{1}{\tau_{\text{non}}}=\frac{4}{9}\frac{n_{\text{out}}}{c^{3}}f_{\varepsilon}^{2}\left(\Egap+\frac{\pi^{2}}{2\mu R^{2}}\right)\Ep\,.\label{eq:tau_non}
\end{equation}

\begin{figure}
\includegraphics[scale=0.54]{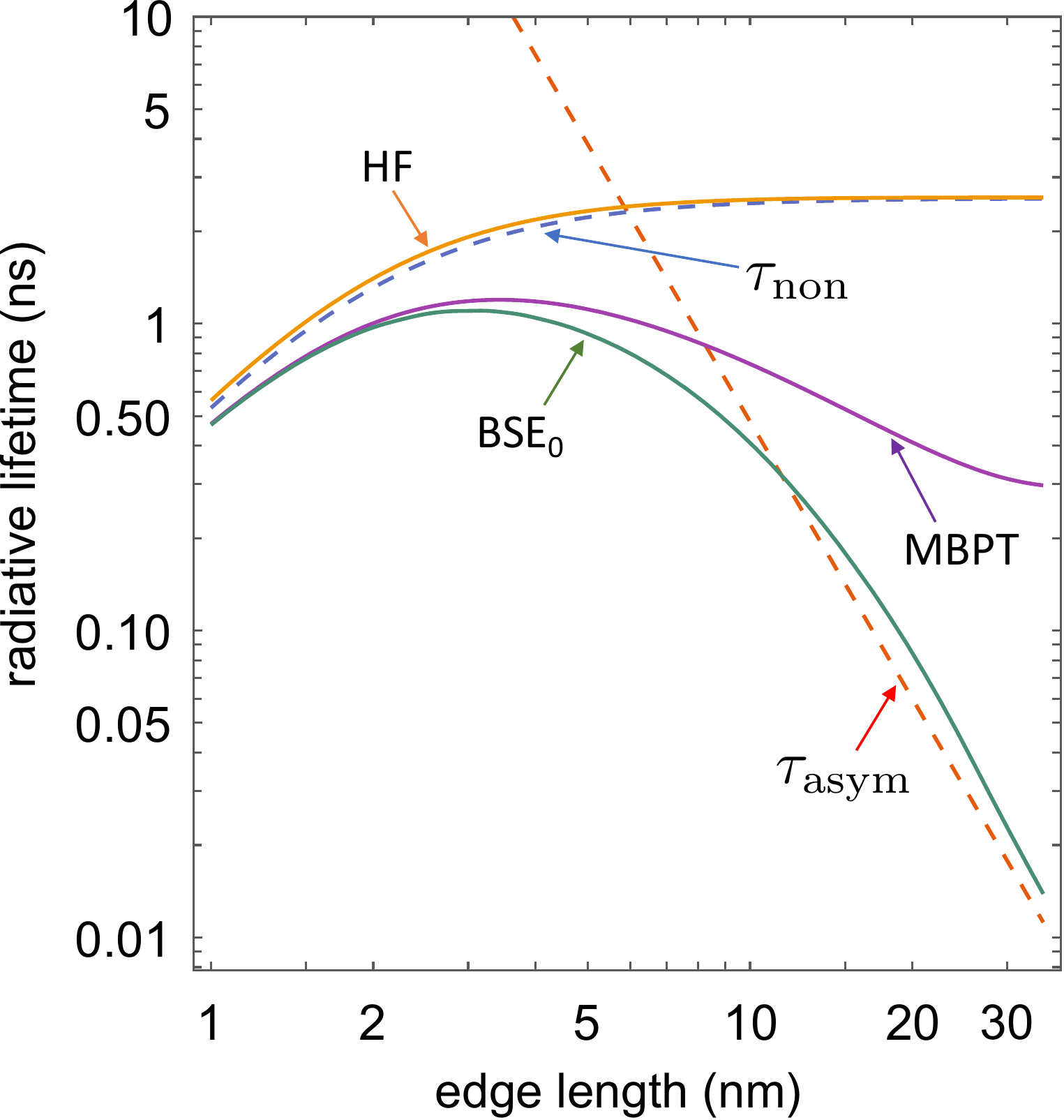}

\caption{\label{fig:tau_rad}Log-log plot of the radiative lifetime of the
$1S_{e}$-$1S_{h}$ ground-state exciton of a NC of CsPbBr$_{3}$
vs.\ NC edge length, in various approximations. All theories assume
the EMA. Dashed lines: $\tau_{\text{asym}}$, large-$R$ asymptotic
form of the lifetime {[}Eq.~(\ref{eq:tau_asym}){]}; $\tau_{\text{non}}$,
lifetime for noninteracting particles {[}Eq.~(\ref{eq:tau_non}){]}.
Continuous lines (top to bottom): HF, configuration-averaged Hartree-Fock;
MBPT, many-body perturbation theory up to first order (following Ref.~\protect\cite{nguyen-20b-psk});
BSE$_{0}$, particle-hole Bethe-Salpeter equation.}
\end{figure}

The other case is the large-$R$ limit for interacting carriers. Here
one makes use of the bound-exciton wave function $\Psi(\mathbf{r}_{e},\mathbf{r}_{h})$~(\ref{eq:2body_soln})
and generalizes the overlap $\BraKet{1S_{e}}{1S_{h}}$ in Eq.~(\ref{eq:redmxel_non})
to~\cite{efros-82-sqd} 
\begin{align}
 & \int\Psi(\mathbf{r}_{e},\mathbf{r}_{h})\delta^{3}(\mathbf{r}_{e}-\mathbf{r}_{h})\,d^{3}\mathbf{r}_{e}d^{3}\mathbf{r}_{h}\nonumber \\
 & \quad\quad\quad\quad=\psi_{\text{rel}}(\bm{0})\int\psi_{\text{CM}}(\mathbf{r}')\,d^{3}\mathbf{r}'\nonumber \\
 & \quad\quad\quad\quad=\frac{2\sqrt{2}}{\pi}\left(\frac{\mu R}{\epsin}\right)^{3/2}\,,\label{eq:ovlp_subst}
\end{align}
which gives the large-$R$ asymptotic form of the lifetime~\cite{efros-82-sqd},
\begin{equation}
\frac{1}{\tau_{\text{asym}}}=\frac{4}{9}\frac{n_{\text{out}}}{c^{3}}f_{\varepsilon}^{2}\left(\Egap-\frac{\mu}{2\epsin^{2}}\right)\Ep\left(\frac{8}{\pi^{2}}\frac{\mu^{3}R^{3}}{\epsin^{3}}\right)\,.\label{eq:tau_asym}
\end{equation}
For large $R$, the lifetime thus goes as $\tau\sim1/R^{3}$.

\begin{figure}
\includegraphics[scale=0.5]{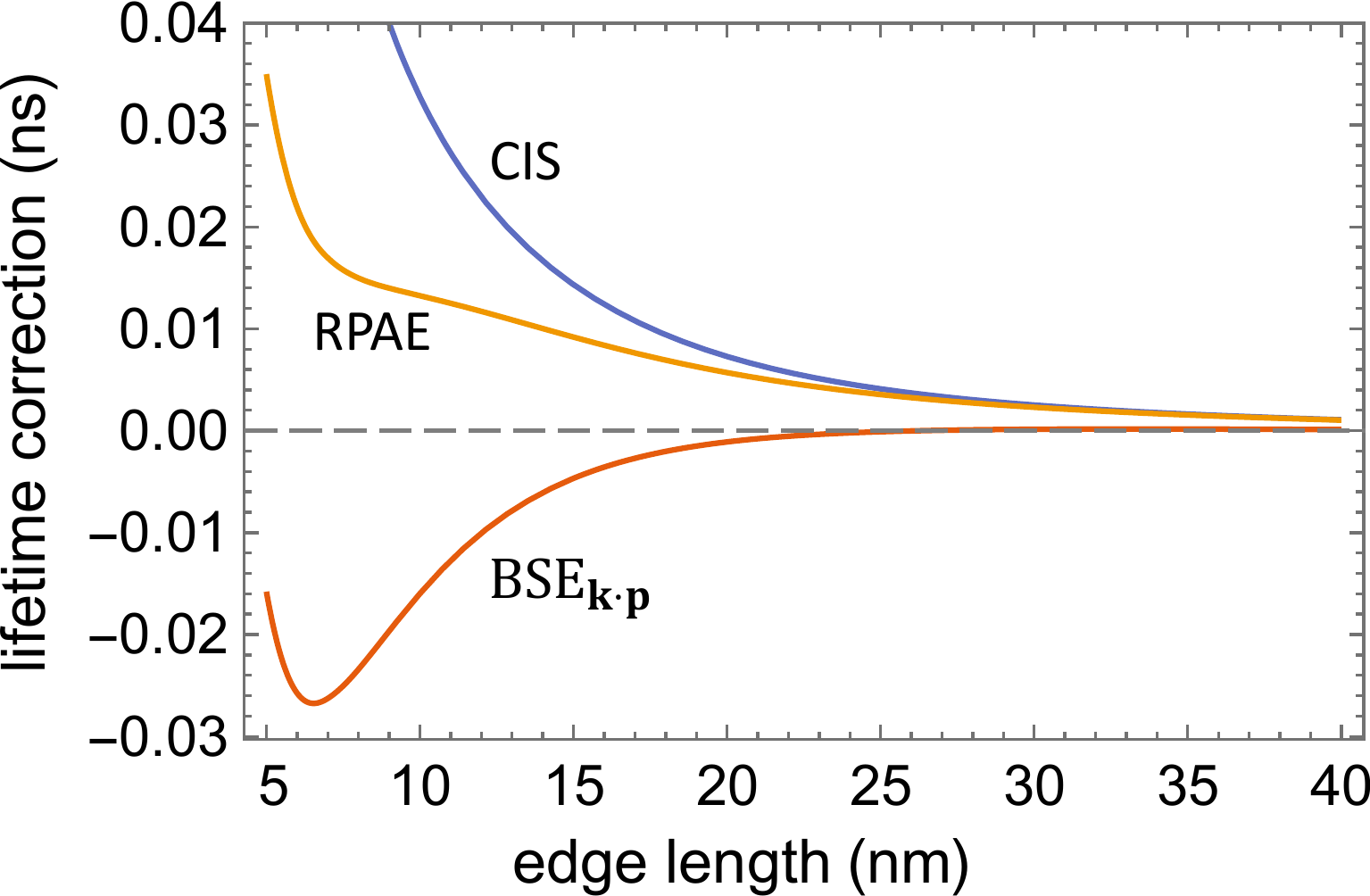}

\caption{\label{fig:tau_kp_corr}Plot of $\kp$ corrections to the lifetime
of the $1S_{e}$-$1S_{h}$ ground-state exciton of a NC of CsPbBr$_{3}$
vs.\ NC edge length, for various levels of theory. The $\kp$ correction
is defined relative to the BSE$_{0}$ value (Fig.~\ref{fig:tau_rad}),
which corresponds to zero. Continuous curves (top to bottom): CIS,
configuration-interaction singles; RPAE, random-phase approximation
with exchange; BSE$_{\kp}$, particle-hole Bethe-Salpeter equation
using basis states of the $4\times4$ $\kp$ model.}
\end{figure}

These analytical results together with various numerical results are
shown in Fig.~\ref{fig:tau_rad} as a function of NC edge length.
One sees that the mean-field HF approach gives almost the same lifetime
as noninteracting carriers, and both tend to a constant as $L\rightarrow\infty$.
Correlation has a large effect as $L$ increases, however. Although
MBPT up to first order~\cite{nguyen-20b-psk} accounts quite well
for the lifetime in the strong-confinement limit $L\ll2a_{B}=6.1$~nm,
it deviates significantly from both BSE$_{0}$ and $\tau_{\text{asym}}$
in intermediate confinement and also has the wrong large-$R$ limit~\cite{nguyen-20b-psk}.
In intermediate confinement, only BSE$_{0}$ gives a satisfactory
description of the radiative lifetime.

\begin{figure}[!b]
\includegraphics[scale=0.5]{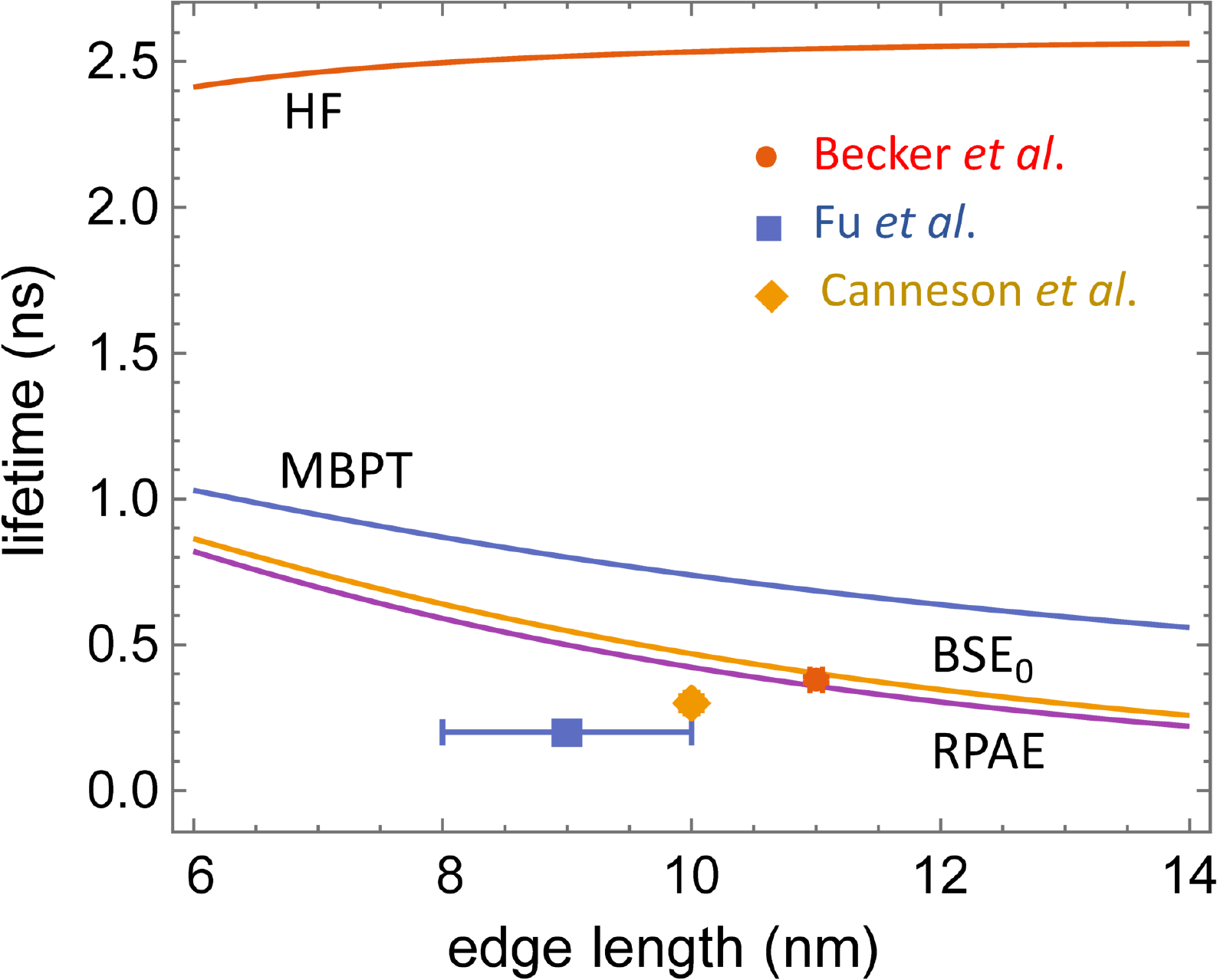}

\caption{\label{fig:tau_expt}Comparison of theoretical and experimental lifetimes
of NCs of CsPbBr$_{3}$. Continuous lines (top to bottom): HF, configuration-averaged
Hartree-Fock (Ref.~\protect\cite{nguyen-20b-psk}); MBPT, many-body
perturbation theory up to first order (Ref.~\protect\cite{nguyen-20b-psk});
BSE$_{0}$, particle-hole Bethe-Salpeter equation using EMA basis
states; RPAE, random-phase approximation with exchange using basis
states of the $4\times4$ $\kp$ model. Experimental values: Becker
\emph{et al}., Ref.~\protect\cite{becker-18-psk}; Fu \emph{et al}.,
Ref.~\protect\cite{fu-17-psk}; Canneson \emph{et al}., Ref.~\protect\cite{canneson-17-psk}.}
\end{figure}

The $\kp$ corrections to the radiative lifetime are shown in Fig.~\ref{fig:tau_kp_corr}
using various approaches within the $4\times4$ $\kp$ model. Overall,
the $\kp$ corrections are small, up to about 5\% of the lifetime
in intermediate confinement. However, the more complete RPAE approach
gives significantly different results from CIS and BSE$_{\kp}$ and,
in situations where $\kp$ corrections are of interest, is therefore
to be preferred.

Our results are compared with the available experimental data (at
cryogenic temperatures) in Fig.~\ref{fig:tau_expt}. In the size
range of interest, the all-order methods give significantly improved
agreement with experiment and bring the theory into good agreement
with the measurement of Becker \emph{et al.}~\cite{becker-18-psk}.
However, the decay rate is approximately proportional to the Kane
parameter $\Ep$ {[}see, e.g., Eq.~(\ref{eq:tau_asym}){]}, whose
value is presently quite uncertain. A larger value of $\Ep$ might
favor the other measurements. We also note that the measurements disagree
with themselves by up to a factor of 2. Further discussion of possible
errors in both theory and experiment is given in Ref.~\cite{nguyen-20b-psk}.

Finally, we note that, in our discussion of the large-$R$ limit of
the lifetime {[}Eq.~(\ref{eq:tau_asym}){]}, we have assumed an idealized
situation in which the carriers remain in a single pure quantum state
(the ground state) for all NC sizes $R$. For sufficiently large sizes
at finite temperature, this assumption will fail and the predicted
decrease in lifetime as $1/R^{3}$ will break down~\cite{takagahara-87-sqd}.
However, Fig.~\ref{fig:tau_expt} makes it clear that a strong renormalization
due to correlation is observable in the experimental data for the
synthesized NC sizes at cryogenic temperatures.

\subsection{\label{subsec:LR-FS}Fine structure: long-range Coulomb interaction}

In this section we illustrate the application of all-order methods
to the ground-state exciton fine structure, focusing on one contribution,
the LR Coulomb interaction~(\ref{eq:g12LR}), which has received
much attention recently~\cite{ben-aich-19-psk,sercel-19-psk,ben-aich-20-psk,tamarat-19-psk}.
Other contributions to the fine structure (with comparable size) include
the SR Coulomb interaction~(\ref{eq:g12SR})~\cite{becker-18-psk,sercel-19-psk,ben-aich-20-psk,tamarat-19-psk},
NC shape and lattice deformations~\cite{ben-aich-19-psk,sercel-19-psk,ben-aich-20-psk},
and a possible strong Rashba interaction~\cite{becker-18-psk,sercel-19-psk,swift-21-psk}.

The leading Coulomb contribution to exciton fine structure is the
exchange interaction, Figs.~\ref{fig:ladders}(c) and (d)~\cite{Knox,pikus-71-sqd}.
As discussed in Sec.~\ref{sec:formalism}, this term is formally
of order $O[(\kp)^{2}]$ in $\kp$ perturbation theory. In the present
approach, $\kp$ corrections enter via the small components of the
wave function in VB-CB-coupled methods such as the $4\times4$ $\kp$
model. By keeping track of both large and small components when evaluating
Coulomb matrix elements, the $\kp$ corrections then propagate ``automatically''
to the final energy. It is thus possible to extract the fine-structure
contribution directly by taking the difference of the total energy
for the $F_{\text{tot}}=1$ and $F_{\text{tot}}=0$ fine-structure
states of the ground-state exciton, testing this difference carefully
for numerical significance.

\begin{figure}
\includegraphics[scale=0.5]{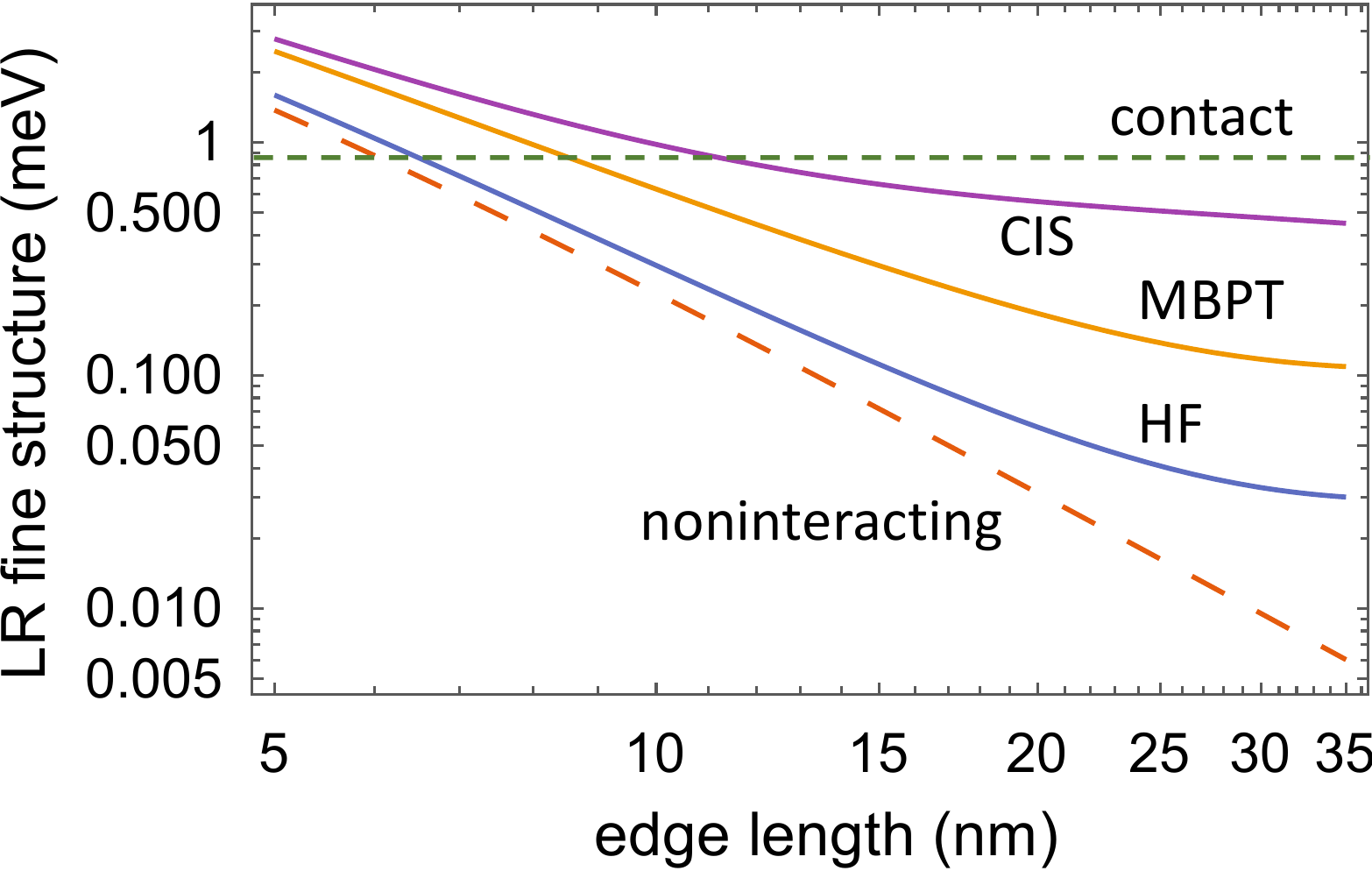}

\caption{\label{fig:LR_fine_structure}Log-log plot of LR fine structure for
a NC of CsPbBr$_{3}$ vs.\ NC edge length, in various approximations.
All approaches use the $4\times4$ $\kp$ model. Dashed lines (top
to bottom): `contact', large-$R$ asymptotic value calculated assuming
an effective contact interaction, Eqs.~(\ref{eq:g12_LRFS}) and (\ref{eq:LRFS_asym});
`noninteracting', noninteracting carriers, Eq.~(\ref{eq:LRFS_non}).
Continuous lines (top to bottom): CIS, configuration-interaction singles;
MBPT, many-body perturbation theory up to first order; HF, configuration-averaged
Hartree-Fock. The curve for RPAE is indistinguishable from that for
CIS, while BSE$_{\kp}$ accounts poorly for the LR fine structure
and is not shown (see text for further discussion).}
\end{figure}

In Fig.~\ref{fig:LR_fine_structure}, we show the LR fine-structure
contribution calculated this way in various many-body approximations
as a function of NC size. A positive value of the fine structure indicates
that the $F_{\text{tot}}=1$ state is higher in energy than the $F_{\text{tot}}=0$
state. Note that the only all-order method shown in the figure is
CIS. The BSE$_{\kp}$ method does not reproduce the fine structure
accurately. This happens because BSE$_{\kp}$ (like BSE$_{0}$) excludes
the last term in Eq.~(\ref{eq:Hmatrix}) {[}Fig.~\ref{fig:Excitations}(e){]},
which is the term that generates the exchange energy {[}Figs.~\ref{fig:ladders}(c)
and (d){]}. The RPAE method does include this term, but the additional
fine-structure $\kp$ corrections contained in RPAE relative to CIS
are very small, giving a modification of only about 1\% or less of
the CIS result over the range of sizes shown. As with the radiative
lifetime, correlation is seen to be a very important effect in intermediate
confinement, and the all-order CIS or RPAE methods give significantly
different results from HF and MBPT in this regime.

As before, it is possible to make analytical progress in two cases
within the EMA. Using external-leg wave-function corrections to first
order in $\kp$ perturbation theory, it can be shown~\cite{pikus-71-sqd,takagahara-93-sqd,tong-11-sqd}
that the lowest-order exchange energy, Fig.~\ref{fig:ladders}(c),
is given in a spherical approximation by the matrix element of the
effective contact interaction,
\begin{equation}
\tilde{g}_{12}=\frac{4\pi}{3}\frac{1}{\epsin\omega_{\alpha}^{2}}\delta^{3}(\mathbf{r}_{1}-\mathbf{r}_{2})\mathbf{p}_{1}\cdot\mathbf{p}_{2}\,.\label{eq:g12_LRFS}
\end{equation}
Here, $\omega_{\alpha}$ is the exciton energy, $\delta^{3}(\mathbf{r}_{1}-\mathbf{r}_{2})$
acts only on the envelope wave functions, and $\mathbf{p}_{1}\cdot\mathbf{p}_{2}$
acts only on the Bloch functions. The operator~(\ref{eq:g12_LRFS})
is strictly valid only when the external states in Fig.~\ref{fig:ladders}(c)
are $S$ states. For external states of higher angular momentum, a
second term with quadrupole symmetry in principle also contributes~\cite{tong-11-sqd}.

Evaluating the matrix element $\BraOperKet{eh}{\tilde{g}_{12}}{he}$
using the noninteracting $1S_{e}$ and $1S_{h}$ wave functions~(\ref{eq:gs_wave_function})
gives the LR fine structure for noninteracting particles~\cite{tamarat-19-psk,sercel-19-psk}
\begin{equation}
\mathcal{F}_{\text{non}}=\frac{4\pi}{9}\frac{\Ep}{\epsin}\left(\Egap+\frac{\pi^{2}}{2\mu R^{2}}\right)^{-2}\frac{\xi}{R^{3}}\,,\label{eq:LRFS_non}
\end{equation}
where $\xi\approx0.672$1. We can use this expression to test our
numerics, by comparing $\mathcal{F}_{\text{non}}$ with the fine structure
calculated numerically from the lowest-order exchange diagram, Fig.~\ref{fig:ladders}(c),
using noninteracting $1S$ states on the external legs that were generated
numerically in the $4\times4$ $\kp$ model (including large and small
components). We find agreement with Eq.~(\ref{eq:LRFS_non}) to about
one part in $10^{4}$ for large $R$. This exercise fails when the
external states have orbital angular momentum greater than zero, because
the effective operator in Eq.~(\ref{eq:g12_LRFS}) excludes the quadrupole
term.

An estimate of the LR fine structure in the large-$R$ limit can also
be made, by evaluating the expectation value $\BraOperKet{\Psi}{\tilde{g}_{12}}{\Psi}$
using the bound-exciton wave function $\Psi$ in Eq.~(\ref{eq:2body_soln}).
This yields
\begin{equation}
\mathcal{F}_{\text{asym}}=\frac{4\pi}{9}\frac{\Ep}{\epsin}\left(\Egap-\frac{\mu}{2\epsin^{2}}\right)^{-2}\left(\frac{1}{\pi}\frac{\mu^{3}}{\epsin^{3}}\right)\,.\label{eq:LRFS_asym}
\end{equation}
$\mathcal{F}_{\text{asym}}$ is a constant, representing the LR fine-structure
contribution of the bulk exciton, with a value $\mathcal{F}_{\text{asym}}=0.869$~meV
for the parameters in Table~\ref{tab:parameters}. However, Eq.~(\ref{eq:LRFS_asym})
is only approximate, because the derivation implicitly assumes external
states with angular momentum greater than zero on the external legs
of the effective operator $\tilde{g}_{12}$~(\ref{eq:g12_LRFS}).
This can be seen by expressing the diagram in Fig.~\ref{fig:ladders}(d)
in terms of all-order CI amplitudes~(\ref{eq:CIexciton}), 
\begin{equation}
\mathcal{F}_{\alpha}=\sum_{ehe'h'}(\mathcal{X}_{e'h'}^{\alpha})^{*}\mathcal{X}_{eh}^{\alpha}\BraOperKet{e'h}{g_{12}}{h'e}\,.\label{eq:LRFS_all-order}
\end{equation}
The sum here is over all exciton channels $(e,h)$ and $(e',h')$,
implying contributions from $P$, $D$, etc., states in the Coulomb
matrix element.

A numerical estimate of the large-$R$ limit can instead be made using
the all-order CIS method, which is set up to handle the angular couplings
in full generality. From Fig.~\ref{fig:LR_fine_structure}, one sees
that the LR fine structure is not quite asymptotic (constant) at $L=35$~nm
and has a numerical value that is of the same order of magnitude as
$\mathcal{F}_{\text{asym}}$, but about half the size.

As mentioned in Sec.~\ref{sec:formalism}, the contribution of the
SR Coulomb interaction to the exciton fine structure is also formally
of order $O[(\kp)^{2}]$, and studies have shown that the SR and LR
fine-structure contributions are comparable \cite{sercel-19-psk,tamarat-19-psk,ben-aich-19-psk}.
The leading SR fine-structure term can be included in the CIS and
RPAE approaches by putting $g_{12}=g_{12}^{\text{LR}}+g_{12}^{\text{SR}}$
in the exchange term $\BraOperKet{eh'}{g_{12}}{he'}$ {[}Fig.~\ref{fig:Excitations}(e){]}
in Eq.~(\ref{eq:Hmatrix}). The $O(1)$ particle-hole ladder terms,
generated by iterating Fig.~\ref{fig:Excitations}(d), will then
provide the vertex-renormalization terms in Fig.~\ref{fig:ladders}(d),
which are also important for the SR fine structure.

\subsection{\label{subsec:1PA_xsection}One-photon absorption cross section}

The one-photon absorption cross section for laser frequency $\omega$
can be written~\cite{elliott-57-sqd,efros-82-sqd} 
\begin{equation}
\sigma^{(1)}(\omega)=\sum_{\alpha}T_{\alpha}^{(1)}\Delta_{\alpha}(\omega-\omega_{\alpha})\,,\label{eq:sigma_1PA}
\end{equation}
where $T_{\alpha}^{(1)}$ is the one-photon ``transition strength''
to a final-state exciton $\alpha$ with energy $\omega_{\alpha}$,
and $\Delta_{\alpha}(\omega-\omega_{\alpha})$ is a line-shape function
for this transition. The transition strength is given by 
\begin{equation}
T_{\alpha}^{(1)}=\frac{4\pi^{2}}{3}\frac{f_{\varepsilon}^{2}}{n_{\text{out}}c\omega}\left|M_{\alpha}\right|^{2}\,,\label{eq:T1_alpha}
\end{equation}
involving the same quantities as Eq.~(\ref{eq:tau_alpha}).

\begin{figure}
\includegraphics[scale=0.60]{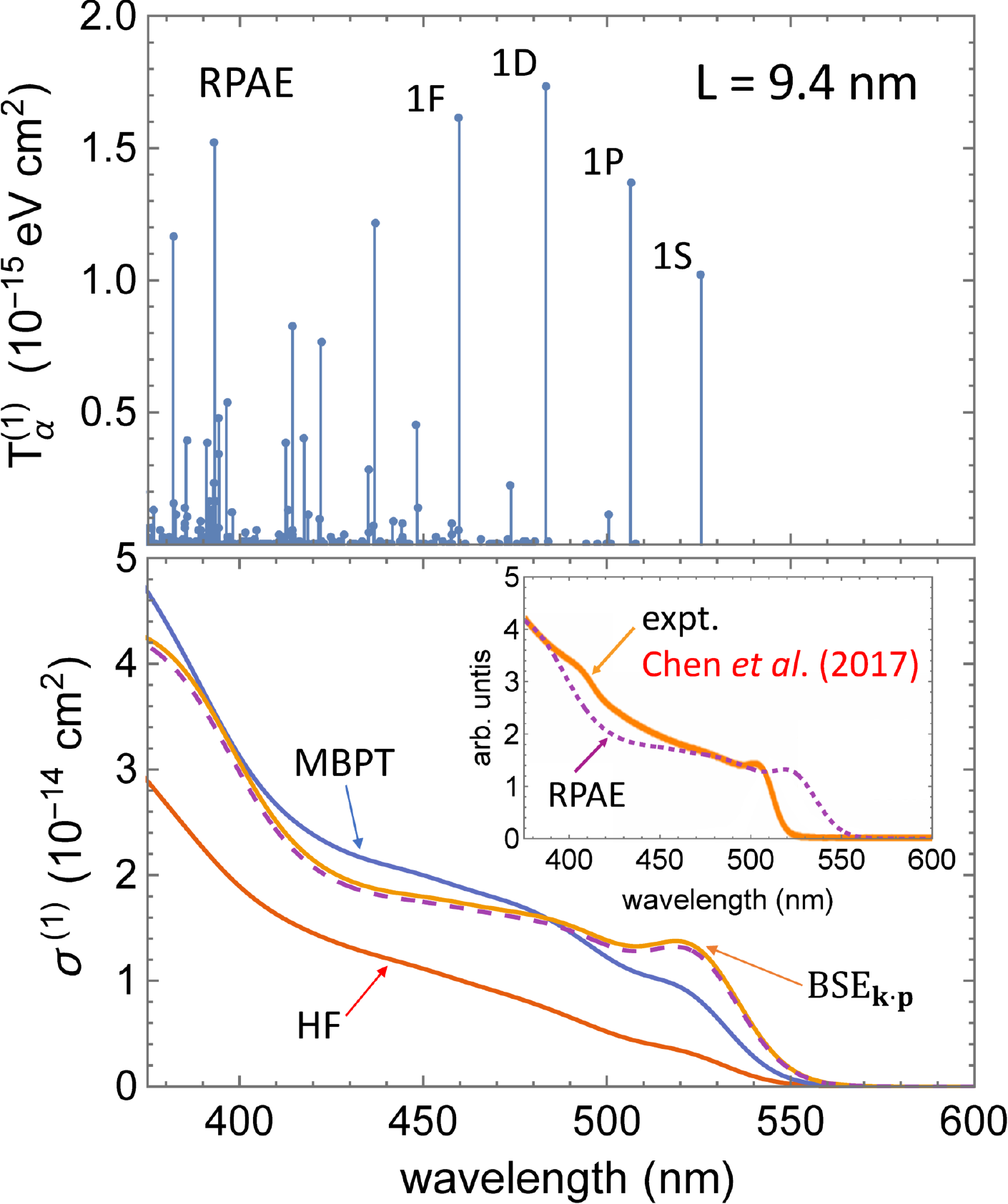}

\caption{\label{fig:1PA}One-photon absorption for a NC of CsPbBr$_{3}$ with
edge length $L=9.4$~nm. \emph{Upper panel}: Transition strengths
$T_{\alpha}^{(1)}$, Eq.~(\ref{eq:T1_alpha}), calculated within
RPAE using the $8\times8$ $\kp$ model. The quantum numbers of some
dominant transitions are indicated; $1S$ indicates a final-state
exciton $1S_{e}$-$1S_{h}$, $1P$ indicates $1P_{e}$-$1P_{h}$,
etc. \emph{Lower panel}: continuous lines show one-photon absorption
cross sections $\sigma^{(1)}$ calculated in various approximations.
HF: configuration-averaged Hartree-Fock; MBPT: many-body perturbation
theory up to first order (following Ref.~\protect\cite{nguyen-20b-psk});
BSE$_{\kp}$: particle-hole Bethe-Salpeter approach using single-particle
states from the $8\times8$ $\kp$ model. The dashed line, which is
nearly coincident with the BSE$_{\kp}$ line, indicates the RPAE approximation.
\emph{Inset of lower panel}: unnormalized one-photon absorption cross
section (arbitrary units) taken from the experiment of Chen \emph{et
al}.~\cite{chen-17-psk}\ for a NC of edge length $L=9.4$~nm.
The experimental absorption cross section is scaled so that the threshold
peak ($1S_{e}$-$1S_{h}$) has the same numerical value as the theoretical
BSE$_{\kp}$ curve.}
\end{figure}

To evaluate Eq.~(\ref{eq:sigma_1PA}), we extract a large number
of correlated exciton states $\alpha$ using the CIS~(\ref{eq:CIeigenvalue})
or RPAE~(\ref{eq:RPAEeigenvalue}) eigenvalue equations, from the
ground state up to a high energy cutoff (typically a few hundred eigenstates
up to an excitation energy of several eV). The computed transition
strengths $T_{\alpha}^{(1)}$ are then broadened using the phenomenological
approach of Ref.~\cite{nguyen-20b-psk} (which should be consulted
for more details). A Gaussian line-shape function is chosen emphasizing
inhomogeneous broadening mechanisms,

\begin{equation}
\Delta_{\alpha}(\omega-\omega_{\alpha})=\frac{1}{\sigma_{\alpha}\sqrt{2\pi}}\exp\left[-\frac{(\omega-\omega_{\alpha})^{2}}{2\sigma_{\alpha}^{2}}\right]\,,\label{eq:line-shape}
\end{equation}
with width parameters 
\begin{equation}
\sigma_{\alpha}^{2}=\left(\sigma_{\alpha}^{\text{size}}\right)^{2}+\left(\sigma_{\alpha}^{\text{other}}\right)^{2}\,.\label{eq:line_width}
\end{equation}
The width $\sigma_{\alpha}^{\text{size}}$ is a transition-dependent
term representing the range of NC sizes present in the ensemble, which
we take as $\delta L/L\approx5$\% for CsPbBr$_{3}$~\cite{chen-17-psk}.
Other broadening mechanisms are represented by $\sigma_{\alpha}^{\text{other}}$,
which is given the constant value 80~meV for all transitions except
the ground-state exciton, for which we take $\sigma_{\alpha}^{\text{other}}=52$~meV.
Because the line-shape functions are normalized to unity, $\int_{0}^{\infty}\Delta_{\alpha}(\omega-\omega_{\alpha})\,d\omega=1$,
the broadening parameters do not change significantly the average
value of the computed cross section $\sigma^{(1)}$, only the appearance
of substructure related to individual transitions $T_{\alpha}^{(1)}$.
The broadening parameters above, which are physically reasonable,
were chosen to reproduce approximately the resolution of individual
transitions observed in the experimental absorption spectra of Ref.~\cite{chen-17-psk}
for an edge length of $L=9.4$~nm (see inset to Fig.~\ref{fig:1PA}).

A study of excitons at cryogenic temperatures in bulk (CH$_{3}$NH$_{3}$)PbBr$_{3}$~\cite{tanaka-03-psk},
which may be expected to have a similar band structure to CsPbBr$_{3}$,
showed a sharp absorption line at an excitation energy of 1.07~eV
above threshold. This was attributed to transitions from the $s$-like
VB ($R_{6}^{+})$ to the $p_{3/2}$-like CB ($R_{8}^{-})$ at the
$R$ point of the Brillouin zone; higher-lying structures around 1.6~eV
above threshold were attributed to transitions to the $p_{1/2}$-like
CB at the $M$ point. The absorption spectra of NCs of CsPbBr$_{3}$~\cite{brennan-17-psk,chen-17-psk}
show analogous features in the form of steps in the absorption cross
section, the first occurring around 0.6--0.8~eV above threshold
(e.g., see the inset to Fig.~\ref{fig:1PA}). Interpreted as the
threshold for absorption to the $p_{3/2}$-like CB band at the $R$
point ($R_{6}^{+}\rightarrow R_{8}^{-}$), this implies a value of
about 0.6--0.8~eV for the spin-orbit coupling parameter $\Esoc$
discussed in Sec.~\ref{subsec:parameters} (with small corrections
due to the confinement shifts present in the NC spectra, which have
an order of magnitude of several tens of meV for an edge length of
9~nm, as shown in Fig.~\ref{fig:conf_energy}). DFT calculations
for CsPbBr$_{3}$~\cite{becker-18-psk,sercel-19-psk} show a similar
energy ordering of band-structure features to that observed for (CH$_{3}$NH$_{3}$)PbBr$_{3}$
in Ref.~\cite{tanaka-03-psk}, although the predicted value of $\Esoc$
is somewhat larger (e.g., $\Esoc=1.54$~eV in Ref.~\cite{sercel-19-psk}).

For this calculation, we use the $8\times8$ $\kp$ model discussed
in Sec.~\ref{subsec:parameters} in order to include the secondary
absorption threshold to the $p_{3/2}$-like CB, assuming $\Esoc=0.8$~eV.
The $8\times8$ $\kp$ model also involves Luttinger parameters $\gamma_{i}$
($i=1$--3) \cite{efros-00-sqd}, which describe the effective-mass
and intraband couplings within the $p_{3/2}$-like CB. The Luttinger
parameters are not known for CsPbBr$_{3}$; we choose $\gamma_{i}$
such that $\tilde{\gamma}_{i}=0$, where $\tilde{\gamma}_{i}$ is
the contribution to $\gamma_{i}$ from remote bands (not included
in the $8\times8$ $\kp$ model). Transition strengths obtained using
RPAE are shown in Fig.~\ref{fig:1PA} (upper panel). The first few
dominant transitions are to excitons of the form $1l_{e}$-$1l_{h}$
(where $l$ denotes orbital angular momentum), which give nonzero
overlaps $\BraKet{1l_{e}}{1l_{h}}$ in Eq.~(\ref{eq:redmxel_non})
for the noninteracting case. However, correlation and $\kp$ corrections
allow numerous other transitions with reduced strength, such as $2S_{e}$-$1S_{h}$,
which add to the overall absorption strength. Also, a line such as
$1P_{e}$-$1P_{h}$ has a small fine structure, with components $1(P_{1/2})_{e}$-$1(P_{1/2})_{h}$,
$1(P_{3/2})_{e}$-$1(P_{1/2})_{h}$, $1(P_{1/2})_{e}$-$1(P_{3/2})_{h}$,
and $1(P_{3/2})_{e}$-$1(P_{3/2})_{h}$.

Theoretical absorption cross sections after line broadening are shown
in the lower panel of Fig.~\ref{fig:1PA} and compared to the experimental
cross section of Chen \emph{et al.}~\cite{chen-17-psk} for a NC
of edge length 9.4~nm. As discussed in Ref.~\cite{nguyen-20b-psk},
measurements of the absolute cross section, which have been performed
in some cases for particular wavelengths, disagree with each other
by up to an order of magnitude. We therefore focus here on the \emph{shape}
of the absorption curve. The all-order cross sections from BSE$_{\kp}$
and RPAE are found to be in good qualitative agreement with the measured
cross section of Ref.~\cite{chen-17-psk}. Similar agreement is found
with Ref.~\cite{brennan-17-psk}, where the absorption curve for
$L=9.2$~nm has a similar shape near threshold, with a step around
0.6~eV above threshold. The threshold peak (at about 510~nm) is
a transition to the $1S_{e}$-$1S_{h}$ exciton. A weaker transition
around 470~nm is just visible in the experimental (and theoretical)
curve, and is due mainly to a combination of the $1P_{e}$-$1P_{h}$,
$1D_{e}$-$1D_{h}$, and $1F_{e}$-$1F_{h}$ transitions. The secondary
transition to the $p_{3/2}$-like CB (around 410~nm) is perhaps more
pronounced in the theoretical curve than in the measurement. The strength
of this transition relative to the threshold transition is fixed in
our approach by the $8\times8$ $\kp$ model, in which the $4\times4$
$\kp$ model is embedded with the ratios of coupling constants constrained
by symmetry. However, the ratio of the cross section at 375~nm to
its value at the threshold peak is in good agreement with experiment.

Note that the theoretical $1S_{e}$-$1S_{h}$ threshold peak, which
is based on a band gap measured at cryogenic temperatures (Table~\ref{tab:parameters}),
is redshifted by about 0.1~eV compared to the experimental peak,
which was measured at room temperature or above. This is likely due
to the temperature-dependent shift of the band gap and possibly also
to a change of crystal phase~\cite{yang-17-psk}. A better fit to
the experiment could have been found by choosing $\Egap$ to be about
0.1~eV larger, compensated by a value of $\Esoc$ that is about 0.1~eV
smaller, so that the step in the absorption curve occurs at the same
wavelength.

\begin{table}
\caption{\label{tab:beta}Transition strengths and renormalization factors
for the principal one-photon-absorption transitions in Fig.~\ref{fig:1PA}.
The columns HF, MBPT (1st-order), and RPAE give the transition strengths
summed over fine-structure components, Eq.~(\ref{eq:T1_summed}),
for the indicated model (units: $10^{-15}$~eV\,cm$^{2}$). The
renormalization factors $\beta_{\text{MBPT}}$ and $\beta_{\text{RPAE}}$
are the enhancement of the transition strength relative to the HF
model.}

\begin{ruledtabular}
\begin{tabular}{lddddd}
Transition & \multicolumn{1}{c}{HF} & \multicolumn{1}{c}{MBPT} & \multicolumn{1}{c}{RPAE} & \multicolumn{1}{c}{$\beta_{\text{MBPT}}$} & \multicolumn{1}{c}{$\beta_{\text{RPAE}}$}\\
\hline 
1S & 0.21 & 0.66 & 1.02 & 3.1 & 4.8\\
1P & 0.58 & 1.46 & 1.38 & 2.5 & 2.4\\
1D & 0.88 & 2.00 & 1.75 & 2.3 & 2.0\\
1F & 1.12 & 2.09 & 1.73 & 1.9 & 1.6\\
\end{tabular}
\end{ruledtabular}

\end{table}

In Fig.~\ref{fig:1PA}, the cross section for HF can be seen to be
qualitatively different, rising steadily from a weak threshold peak.
The cross section obtained using 1st-order MBPT~\cite{nguyen-20b-psk}
is a partial improvement on the HF cross section. To understand this
phenomenon in more detail, we consider the transition strength of
the first few dominant transitions in the various approaches. We find
that the various many-body treatments tend to redistribute the transition
strength among the fine-structure components in different ways. Since
the fine-structure lines are nearly coincident and only the sum of
their transition strengths contributes to the observable cross section,
for this analysis we therefore sum the transition strength over fine-structure
components, 
\begin{equation}
\tilde{T}_{\alpha}^{(1)}=\sum_{F,F'}T_{\alpha}^{(1)}(F,F')\,.\label{eq:T1_summed}
\end{equation}
Summed transition strengths are given in Table~\ref{tab:beta}. We
also give the enhancement or renormalization factor $\beta$ due to
correlation, which is defined as the ratio of $\tilde{T}_{\alpha}^{(1)}$
in a given many-body approach to its value at HF level.

The renormalization factor for the threshold $1S_e$-$1S_{h}$ transition
is large, about 4.8 for $L=9.4$~nm, and increases further (approximately
as $L^{3}$) for larger $L$. This is the same enhancement factor
that applies to the radiative decay rate of the ground-state exciton
discussed in Sec.~\ref{subsec:lifetime}. However, as emphasized
in Ref.~\cite{nguyen-20b-psk}, the renormalization factor decreases
rapidly with increasing energy, approaching unity. In fact, while
1st-order MBPT underestimates the $1S_e$-$1S_{h}$ enhancement factor,
it overestimates slightly the factor for the excited-state transitions.
The effect of this is that the absorption cross section in an all-order
approach tends to have a more prominent threshold peak, followed by
a ``leveling out'' of the cross section at higher energies, bringing
the shape of the absorption cross section into better agreement with
experiment.

\section{\label{sec:Conclusions} Conclusions}

We have presented various many-body formalisms, within a $\kp$ envelope-function
approach, for treating all-order correlated single excitons in NC
quantum dots. These formalisms apply both to the EMA and to $\kp$
models in which the VB and CB are coupled, such as the $4\times4$
$\kp$ model. The latter allow one to treat the ``$\kp$ corrections''
to the EMA. The simplest many-body approach, valid to order $O(1)$
in $\kp$ perturbation theory, was BSE$_{0}$ using an EMA basis set.
The most complete treatment of $\kp$ corrections was given by RPAE.
The various formalisms were explicitly adapted to spherical symmetry
and expressed in terms of radial integrals and angular factors. The
resulting approaches are very rapid and accurate for systems in intermediate
confinement, where correlation effects are in general strong (the
methods typically require a few seconds of computation time on a single
core).

To illustrate these techniques, we considered a class of semiconductor
NCs of great recent interest, inorganic lead-halide perovskites such
as CsPbBr$_{3}$. The most commonly synthesized size range of these
NCs corresponds to intermediate confinement. We showed that in this
size range, the all-order methods gave significant improvements in
accuracy compared to mean-field methods (HF) or to perturbative methods
(MBPT to first or second order), and were significantly more accurate
also than asymptotic results that can be derived analytically for
the infinite-size limit. This was shown to be true for the correlation
energy, the radiative lifetime of the ground-state exciton, and the
LR Coulomb contribution to the exciton fine structure. Partly as a
test of our methods, we also checked that the all-order formalism
had the expected large-size limit, where this was known.

The all-order correlation formalism allows one to generate correlated
excited states rigorously. We used excited states to investigate the
one-photon absorption cross section, where the all-order methods were
shown to give a significant improvement in the shape of the cross
section (versus laser wavelength) near and just above threshold. Also,
because the $\kp$ corrections are integrated directly into the all-order
formalism for VB-CB-coupled models, the approach allows one to calculate
the LR exciton fine structure, an order $O[(\kp)^{2}]$ effect, by
direct subtraction of the total energy of the fine-structure levels.
In other problems, such as the exciton correlation energy or lifetime,
the $\kp$ corrections were found to be quite small (e.g., up to about
5\% for the lifetime). However, it was shown that, in situations where
these $\kp$ corrections are interesting, the more complete RPAE method
can give significantly different $\kp$ corrections than the other
methods, such as BSE$_{\kp}$, and RPAE is therefore to be preferred.

We considered only single excitons in this paper. Other excitonic
systems, such as trions or biexcitons, can be treated by generalized
CI approaches~\cite{shumway-01-sqd,tyrrell-15-sqd} or quantum Monte
Carlo~\cite{shumway-01-sqd}. 
\begin{acknowledgments}
The authors would like to thank T.\ P.\ T.\ Nguyen and Sum T.\ C.\ for
helpful discussions. S.A.B.\ is grateful to Fr\'ed\'eric Schuster
and the CEA's PTC program ``Materials and Processes'' for financial
support. 
\end{acknowledgments}

% APPENDICES

%\medskip{}

\appendix*

\section{\label{app:angular-reduction}Angular reduction}

In this Appendix, we give the reduction of the all-order equations
to radial integrals and angular factors for a spherically symmetric
potential $\Vconf+U$ {[}Eq.~(\ref{eq:basis}){]}. The basis states
$\Ket{e}$ and $\Ket{h}$ have definite total angular momentum $F_{e}$
and $F_{h}$ arising from a coupling of orbital, spin, and Bloch band
angular momenta~\cite{ekimov-93-sqd}; $F_{e}$ and $F_{h}$ couple
in turn to a definite total angular momentum $F_{\text{tot}}$ (and
parity) for each exciton state. (Here, $F_{e}$ and $F_{h}$ are half-integers
and $F_{\text{tot}}$ is an integer.)

It follows that the amplitudes $\mathcal{X}_{eh}^{\alpha}$ and $\mathcal{Y}_{eh}^{\alpha}$
in Eqs.~(\ref{eq:CIeigenvalue}) and (\ref{eq:RPAEeigenvalue}) can
be expressed in terms of Clebsch-Gordon coefficients and \emph{reduced}
amplitudes $\mathcal{\bar{X}}_{eh}^{\alpha}$ and $\mathcal{\bar{Y}}_{eh}^{\alpha}$,
which do not depend on the magnetic substates $m_{e}$ and $m_{h}$
of $e$ and $h$, 
\begin{align}
\mathcal{X}_{eh}^{\alpha} & =\sum_{F_{\text{tot}}M_{\text{tot}}}\mathcal{\bar{X}}_{eh}^{\alpha}(F_{\text{tot}})\nonumber \\
 & \quad\quad\times(-1)^{F_{h}-m_{h}}\BraKet{F_{e}m_{e},F_{h}-m_{h}}{F_{\text{tot}}M_{\text{tot}}}\,,\label{eq:Xreduced}\\
\mathcal{Y}_{eh}^{\alpha} & =\sum_{F_{\text{tot}}M_{\text{tot}}}\mathcal{\bar{Y}}_{eh}^{\alpha}(F_{\text{tot}})\nonumber \\
 & \quad\quad\times(-1)^{F_{e}-m_{e}}\BraKet{F_{e}-m_{e},F_{h}m_{h}}{F_{\text{tot}}M_{\text{tot}}}\,.\label{eq:Yreduced}
\end{align}
The equation for $\mathcal{X}_{eh}^{\alpha}$ corresponds to the coupling
of $e$ and $h$ in Eq.~(\ref{eq:ehexciton}) to a total angular
momentum $F_{\text{tot}}$, summed over all allowed values of $F_{\text{tot}}$,
with the phase factor and minus sign $-m_{h}$ appearing because $h$
is associated with an annihilation operator in Eq.~(\ref{eq:ehexciton})~\cite{Brink-Satchler}.
In the coefficient $\mathcal{Y}_{eh}^{\alpha}$, the roles of $e$
and $h$ are interchanged {[}for example, see Eq.~(\ref{eq:RPAEmxel}){]}.

The Coulomb matrix element can also be expressed in terms of a reduced
two-body matrix element $X_{K}(abcd)$~\cite{Brink-Satchler}, 
\begin{align}
 & \BraOperKet{ab}{g_{12}}{cd}=\sum_{K=0}^{\infty}\sum_{M=-K}^{K}(-1)^{F_{a}+F_{b}+K-m_{a}-m_{b}-M}\nonumber \\
 & \quad\times\Threej{F_{a}}{K}{F_{c}}{-m_{a}}{M}{m_{c}}\Threej{F_{b}}{K}{F_{d}}{-m_{b}}{-M}{m_{d}}X_{K}(abcd)\,,\label{eq:red2body}
\end{align}
where $K$ is a multipole that is in practice limited by angular-momentum
and parity selection rules. In Refs.~\cite{nguyen-20b-psk} and~\cite{nguyen-20c-psk},
expressions are given for $X_{K}(abcd)$ in terms of the radial functions
and the quantum numbers of the states $a$, $b$, $c$, and $d$ in
the $4\times4$ $\kp$ and EMA models.

The RPAE eigenvalue equation for reduced amplitudes $\mathcal{\bar{X}}_{eh}^{\alpha}$
and $\mathcal{\bar{Y}}_{eh}^{\alpha}$ is found by substituting Eqs.~(\ref{eq:Xreduced})--(\ref{eq:red2body})
into Eq.~(\ref{eq:RPAEeigenvalue}) and summing over magnetic substates
using methods of Racah algebra~\cite{Brink-Satchler}. One finds
\begin{equation}
\left(\begin{array}{cc}
\bar{A} & \bar{B}\\
\bar{B} & \bar{A}
\end{array}\right)\left(\begin{array}{c}
\bar{\bm{\mathcal{X}}}^{\alpha}\\
\bar{\bm{\mathcal{Y}}}^{\alpha}
\end{array}\right)=\omega_{\alpha}\left(\begin{array}{cc}
1 & 0\\
0 & -1
\end{array}\right)\left(\begin{array}{c}
\bar{\bm{\mathcal{X}}}^{\alpha}\\
\bar{\bm{\mathcal{Y}}}^{\alpha}
\end{array}\right)\,,\label{eq:RPAEred_eigenvalue}
\end{equation}
where the reduced matrices $\bar{A}$ and $\bar{B}$ are 
\begin{align}
 & \bar{A}_{eh,e'h'}(F_{\text{tot}})=(\epsilon_{e}-\epsilon_{h})\delta_{ee'}\delta_{hh'}\nonumber \\
 & \quad+\BraOperKet{e}{(-U)}{e'}\delta_{hh'}-\BraOperKet{h'}{(-U)}{h}\delta_{ee'}\nonumber \\
 & \quad-(-1)^{F_{\text{tot}}+F_{e'}+F_{h'}}\sum_{K=0}^{\infty}\Sixj{F_{h}}{K}{F_{h'}}{F_{e'}}{F_{\text{tot}}}{F_{e}}X_{K}(eh'e'h)\nonumber \\
 & \quad-(-1)^{F_{\text{tot}}+F_{e'}+F_{h'}}\frac{1}{2F_{\text{tot}}+1}X_{F_{\text{tot}}}(eh'he')\,,\label{eq:Aredmatrix}
\end{align}
and 
\begin{eqnarray}
\bar{B}_{eh,e'h'}(F_{\text{tot}}) & = & -\sum_{K=0}^{\infty}\Sixj{F_{h}}{K}{F_{e'}}{F_{h'}}{F_{\text{tot}}}{F_{e}}X_{K}(ee'h'h)\nonumber \\
 &  & {}-\frac{1}{2F_{\text{tot}}+1}X_{F_{\text{tot}}}(ee'hh')\,.\label{eq:Bredmatrix}
\end{eqnarray}
The equations for different values of $F_{\text{tot}}$ and parity
decouple, giving one eigenvalue equation for each $F_{\text{tot}}$
and parity. A matrix element $\BraOperKet{a}{(-U)}{b}$ in Eq.~(\ref{eq:Aredmatrix})
is the matrix element of a radial potential $-U(r)$, implying that
$F_{a}=F_{b}$ and that the states $a$ and $b$ have the same parity.
Note that the reduced matrices $\bar{A}$ and $\bar{B}$ given by
these equations are real and symmetric.

The RPAE normalization condition (\ref{eq:RPAEnorm}) becomes 
\begin{equation}
\sum_{eh}\left[|\bar{\mathcal{X}}_{eh}^{\alpha}(F_{\text{tot}})|^{2}-|\bar{\mathcal{Y}}_{eh}^{\alpha}(F_{\text{tot}})|^{2}\right]=1\,,\label{eq:RPAErednorm}
\end{equation}
and the reduced matrix element of a one-body operator $M$ with spherical
tensor rank $\kappa$ is given by 
\begin{align}
 & \RME{\alpha(F_{\text{tot}})}{M}{0}=\delta(F_{\text{tot}},\kappa)\sum_{eh}\left[\bar{\mathcal{X}}_{eh}^{\alpha}(F_{\text{tot}})\RME{e}{M}{h}\right.\nonumber \\
 & \quad\quad\quad\quad\left.{}+(-1)^{F_{e}+F_{h}+\kappa}\,\bar{\mathcal{Y}}_{eh}^{\alpha}(F_{\text{tot}})\RME{h}{M}{e}\right]\,.\label{eq:RPAEredmxel}
\end{align}
An important case is the momentum operator, $M=p$, which arises in
interband absorption and emission~\cite{Kira-Koch}.
Expressions for $\RME{a}{p}{b}$ in the $4\times4$ $\kp$ model (and
in the EMA) are given in Appendix A of Ref.~\cite{nguyen-20b-psk},
in the form of radial integrals and angular factors, and including
all $\kp$ corrections arising from the small and large components
of the wave functions. In one-photon absorption, the selection rule
in Eq.~(\ref{eq:RPAEredmxel}) becomes $\delta(F_{\text{tot}},1)$,
corresponding to the conservation of angular momentum for absorption
of a photon from the NC ground state.

In the CIS approach, only the $\bar{A}$ matrix enters and the reduced
eigenvalue equation becomes 
\begin{equation}
\bar{A}\,\bar{\bm{\mathcal{X}}}^{\alpha}=\omega_{\alpha}\bar{\bm{\mathcal{X}}}^{\alpha}\,.\label{eq:CIredeigenvalue}
\end{equation}
The BSE formalism is obtained as usual from the CIS formalism by dropping
the last term in Eq.~(\ref{eq:Aredmatrix}). In CIS and BSE, the
normalization and reduced matrix elements are given by Eqs.~(\ref{eq:RPAErednorm})
and (\ref{eq:RPAEredmxel}), respectively, with $\bar{\mathcal{Y}}_{eh}^{\alpha}=0$.

% BIBLIOGRAPHY



\begin{thebibliography}{47}%
\makeatletter
\providecommand \@ifxundefined [1]{%
 \@ifx{#1\undefined}
}%
\providecommand \@ifnum [1]{%
 \ifnum #1\expandafter \@firstoftwo
 \else \expandafter \@secondoftwo
 \fi
}%
\providecommand \@ifx [1]{%
 \ifx #1\expandafter \@firstoftwo
 \else \expandafter \@secondoftwo
 \fi
}%
\providecommand \natexlab [1]{#1}%
\providecommand \enquote  [1]{``#1''}%
\providecommand \bibnamefont  [1]{#1}%
\providecommand \bibfnamefont [1]{#1}%
\providecommand \citenamefont [1]{#1}%
\providecommand \href@noop [0]{\@secondoftwo}%
\providecommand \href [0]{\begingroup \@sanitize@url \@href}%
\providecommand \@href[1]{\@@startlink{#1}\@@href}%
\providecommand \@@href[1]{\endgroup#1\@@endlink}%
\providecommand \@sanitize@url [0]{\catcode `\\12\catcode `\$12\catcode
  `\&12\catcode `\#12\catcode `\^12\catcode `\_12\catcode `\%12\relax}%
\providecommand \@@startlink[1]{}%
\providecommand \@@endlink[0]{}%
\providecommand \url  [0]{\begingroup\@sanitize@url \@url }%
\providecommand \@url [1]{\endgroup\@href {#1}{\urlprefix }}%
\providecommand \urlprefix  [0]{URL }%
\providecommand \Eprint [0]{\href }%
\providecommand \doibase [0]{https://doi.org/}%
\providecommand \selectlanguage [0]{\@gobble}%
\providecommand \bibinfo  [0]{\@secondoftwo}%
\providecommand \bibfield  [0]{\@secondoftwo}%
\providecommand \translation [1]{[#1]}%
\providecommand \BibitemOpen [0]{}%
\providecommand \bibitemStop [0]{}%
\providecommand \bibitemNoStop [0]{.\EOS\space}%
\providecommand \EOS [0]{\spacefactor3000\relax}%
\providecommand \BibitemShut  [1]{\csname bibitem#1\endcsname}%
\let\auto@bib@innerbib\@empty
%</preamble>
\bibitem [{\citenamefont {Protesescu}\ \emph {et~al.}(2015)\citenamefont
  {Protesescu}, \citenamefont {Yakunin}, \citenamefont {Bodnarchuk},
  \citenamefont {Krieg}, \citenamefont {Caputo}, \citenamefont {Hendon},
  \citenamefont {Yang}, \citenamefont {Walsh},\ and\ \citenamefont
  {Kovalenko}}]{protesescu-15-psk}%
  \BibitemOpen
  \bibfield  {author} {\bibinfo {author} {\bibfnamefont {L.}~\bibnamefont
  {Protesescu}}, \bibinfo {author} {\bibfnamefont {S.}~\bibnamefont {Yakunin}},
  \bibinfo {author} {\bibfnamefont {M.~I.}\ \bibnamefont {Bodnarchuk}},
  \bibinfo {author} {\bibfnamefont {F.}~\bibnamefont {Krieg}}, \bibinfo
  {author} {\bibfnamefont {R.}~\bibnamefont {Caputo}}, \bibinfo {author}
  {\bibfnamefont {C.~H.}\ \bibnamefont {Hendon}}, \bibinfo {author}
  {\bibfnamefont {R.~X.}\ \bibnamefont {Yang}}, \bibinfo {author}
  {\bibfnamefont {A.}~\bibnamefont {Walsh}},\ and\ \bibinfo {author}
  {\bibfnamefont {M.~V.}\ \bibnamefont {Kovalenko}},\ }\bibfield  {title}
  {\bibinfo {title} {Nanocrystals of cesium lead halide perovskites
  ({CsPbX$_{3}$}, {X = Cl, Br, and I}): Novel optoelectronic materials showing
  bright emission with wide color gamut},\ }\href
  {https://pubs.acs.org/doi/10.1021/nl5048779} {\bibfield  {journal} {\bibinfo
  {journal} {Nano Lett.}\ }\textbf {\bibinfo {volume} {15}},\ \bibinfo {pages}
  {3692} (\bibinfo {year} {2015})}\BibitemShut {NoStop}%
\bibitem [{\citenamefont {Becker}\ \emph {et~al.}(2018)\citenamefont {Becker},
  \citenamefont {Vaxenburg}, \citenamefont {Nedelcu}, \citenamefont {Sercel},
  \citenamefont {Shabaev}, \citenamefont {Mehl}, \citenamefont {Michopoulos},
  \citenamefont {Lambrakos}, \citenamefont {Bernstein}, \citenamefont {Lyons},
  \citenamefont {St{\"o}ferle}, \citenamefont {Mahrt}, \citenamefont
  {Kovalenko}, \citenamefont {Norris}, \citenamefont {Rain{\`o}},\ and\
  \citenamefont {Efros}}]{becker-18-psk}%
  \BibitemOpen
  \bibfield  {author} {\bibinfo {author} {\bibfnamefont {M.~A.}\ \bibnamefont
  {Becker}}, \bibinfo {author} {\bibfnamefont {R.}~\bibnamefont {Vaxenburg}},
  \bibinfo {author} {\bibfnamefont {G.}~\bibnamefont {Nedelcu}}, \bibinfo
  {author} {\bibfnamefont {P.~C.}\ \bibnamefont {Sercel}}, \bibinfo {author}
  {\bibfnamefont {A.}~\bibnamefont {Shabaev}}, \bibinfo {author} {\bibfnamefont
  {M.~J.}\ \bibnamefont {Mehl}}, \bibinfo {author} {\bibfnamefont {J.~G.}\
  \bibnamefont {Michopoulos}}, \bibinfo {author} {\bibfnamefont {S.~G.}\
  \bibnamefont {Lambrakos}}, \bibinfo {author} {\bibfnamefont {N.}~\bibnamefont
  {Bernstein}}, \bibinfo {author} {\bibfnamefont {J.~L.}\ \bibnamefont
  {Lyons}}, \bibinfo {author} {\bibfnamefont {T.}~\bibnamefont {St{\"o}ferle}},
  \bibinfo {author} {\bibfnamefont {R.~F.}\ \bibnamefont {Mahrt}}, \bibinfo
  {author} {\bibfnamefont {M.~V.}\ \bibnamefont {Kovalenko}}, \bibinfo {author}
  {\bibfnamefont {D.~J.}\ \bibnamefont {Norris}}, \bibinfo {author}
  {\bibfnamefont {G.}~\bibnamefont {Rain{\`o}}},\ and\ \bibinfo {author}
  {\bibfnamefont {A.~L.}\ \bibnamefont {Efros}},\ }\bibfield  {title} {\bibinfo
  {title} {Bright triplet excitons in caesium lead halide perovskites},\ }\href
  {http://dx.doi.org/10.1038/nature25147} {\bibfield  {journal} {\bibinfo
  {journal} {Nature}\ }\textbf {\bibinfo {volume} {553}},\ \bibinfo {pages}
  {189} (\bibinfo {year} {2018})}\BibitemShut {NoStop}%
\bibitem [{\citenamefont {Chaudhary}\ \emph {et~al.}(2021)\citenamefont
  {Chaudhary}, \citenamefont {Kshetri}, \citenamefont {Kim}, \citenamefont
  {Lee},\ and\ \citenamefont {Kim}}]{chaudhary-21-psk}%
  \BibitemOpen
  \bibfield  {author} {\bibinfo {author} {\bibfnamefont {B.}~\bibnamefont
  {Chaudhary}}, \bibinfo {author} {\bibfnamefont {Y.~K.}\ \bibnamefont
  {Kshetri}}, \bibinfo {author} {\bibfnamefont {H.~S.}\ \bibnamefont {Kim}},
  \bibinfo {author} {\bibfnamefont {S.~W.}\ \bibnamefont {Lee}},\ and\ \bibinfo
  {author} {\bibfnamefont {T.~H.}\ \bibnamefont {Kim}},\ }\bibfield  {title}
  {\bibinfo {title} {Current status on synthesis, properties and applications
  of {CsPbX3 (X = Cl, Br, I)} perovskite quantum dots/nanocrystals},\ }\href
  {https://doi.org/10.1088/1361-6528/ac2537} {\bibfield  {journal} {\bibinfo
  {journal} {Nanotechnology}\ }\textbf {\bibinfo {volume} {32}},\ \bibinfo
  {pages} {502007} (\bibinfo {year} {2021})}\BibitemShut {NoStop}%
\bibitem [{\citenamefont {Li}\ \emph {et~al.}(2016)\citenamefont {Li},
  \citenamefont {Rivarola}, \citenamefont {Davis}, \citenamefont {Bai},
  \citenamefont {Jellicoe}, \citenamefont {de~la Pena}, \citenamefont {Hou},
  \citenamefont {Ducati}, \citenamefont {Gao}, \citenamefont {Friend},
  \citenamefont {Greenham},\ and\ \citenamefont {Tan}}]{li-16-psk}%
  \BibitemOpen
  \bibfield  {author} {\bibinfo {author} {\bibfnamefont {G.~R.}\ \bibnamefont
  {Li}}, \bibinfo {author} {\bibfnamefont {F.~W.~R.}\ \bibnamefont {Rivarola}},
  \bibinfo {author} {\bibfnamefont {N.~J. L.~K.}\ \bibnamefont {Davis}},
  \bibinfo {author} {\bibfnamefont {S.}~\bibnamefont {Bai}}, \bibinfo {author}
  {\bibfnamefont {T.~C.}\ \bibnamefont {Jellicoe}}, \bibinfo {author}
  {\bibfnamefont {F.}~\bibnamefont {de~la Pena}}, \bibinfo {author}
  {\bibfnamefont {S.~C.}\ \bibnamefont {Hou}}, \bibinfo {author} {\bibfnamefont
  {C.}~\bibnamefont {Ducati}}, \bibinfo {author} {\bibfnamefont
  {F.}~\bibnamefont {Gao}}, \bibinfo {author} {\bibfnamefont {R.~H.}\
  \bibnamefont {Friend}}, \bibinfo {author} {\bibfnamefont {N.~C.}\
  \bibnamefont {Greenham}},\ and\ \bibinfo {author} {\bibfnamefont {Z.~K.}\
  \bibnamefont {Tan}},\ }\bibfield  {title} {\bibinfo {title} {Highly efficient
  perovskite nanocrystal light-emitting diodes enabled by a universal
  crosslinking method},\ }\href {https://doi.org/10.1002/adma.201600064}
  {\bibfield  {journal} {\bibinfo  {journal} {Adv.\ Mater.}\ }\textbf {\bibinfo
  {volume} {28}},\ \bibinfo {pages} {3528} (\bibinfo {year}
  {2016})}\BibitemShut {NoStop}%
\bibitem [{\citenamefont {Deng}\ \emph {et~al.}(2016)\citenamefont {Deng},
  \citenamefont {Xu}, \citenamefont {Zhang}, \citenamefont {Zhang},
  \citenamefont {Jin}, \citenamefont {Wang}, \citenamefont {Lee},\ and\
  \citenamefont {Jie}}]{deng-16-psk}%
  \BibitemOpen
  \bibfield  {author} {\bibinfo {author} {\bibfnamefont {W.}~\bibnamefont
  {Deng}}, \bibinfo {author} {\bibfnamefont {X.~Z.}\ \bibnamefont {Xu}},
  \bibinfo {author} {\bibfnamefont {X.~J.}\ \bibnamefont {Zhang}}, \bibinfo
  {author} {\bibfnamefont {Y.~D.}\ \bibnamefont {Zhang}}, \bibinfo {author}
  {\bibfnamefont {X.~C.}\ \bibnamefont {Jin}}, \bibinfo {author} {\bibfnamefont
  {L.}~\bibnamefont {Wang}}, \bibinfo {author} {\bibfnamefont {S.~T.}\
  \bibnamefont {Lee}},\ and\ \bibinfo {author} {\bibfnamefont {J.~S.}\
  \bibnamefont {Jie}},\ }\bibfield  {title} {\bibinfo {title} {Organometal
  halide perovskite quantum dot light-emitting diodes},\ }\href
  {https://doi.org/10.1002/adfm.201601054} {\bibfield  {journal} {\bibinfo
  {journal} {Adv.\ Funct.\ Mater.}\ }\textbf {\bibinfo {volume} {26}},\
  \bibinfo {pages} {4797} (\bibinfo {year} {2016})}\BibitemShut {NoStop}%
\bibitem [{\citenamefont {Yakunin}\ \emph {et~al.}(2015)\citenamefont
  {Yakunin}, \citenamefont {Protesescu}, \citenamefont {Krieg}, \citenamefont
  {Bodnarchuk}, \citenamefont {Nedelcu}, \citenamefont {Humer}, \citenamefont
  {De~Luca}, \citenamefont {Fiebig}, \citenamefont {Heiss},\ and\ \citenamefont
  {Kovalenko}}]{yakunin-15-psk}%
  \BibitemOpen
  \bibfield  {author} {\bibinfo {author} {\bibfnamefont {S.}~\bibnamefont
  {Yakunin}}, \bibinfo {author} {\bibfnamefont {L.}~\bibnamefont {Protesescu}},
  \bibinfo {author} {\bibfnamefont {F.}~\bibnamefont {Krieg}}, \bibinfo
  {author} {\bibfnamefont {M.~I.}\ \bibnamefont {Bodnarchuk}}, \bibinfo
  {author} {\bibfnamefont {G.}~\bibnamefont {Nedelcu}}, \bibinfo {author}
  {\bibfnamefont {M.}~\bibnamefont {Humer}}, \bibinfo {author} {\bibfnamefont
  {G.}~\bibnamefont {De~Luca}}, \bibinfo {author} {\bibfnamefont
  {M.}~\bibnamefont {Fiebig}}, \bibinfo {author} {\bibfnamefont
  {W.}~\bibnamefont {Heiss}},\ and\ \bibinfo {author} {\bibfnamefont {M.~V.}\
  \bibnamefont {Kovalenko}},\ }\bibfield  {title} {\bibinfo {title}
  {Low-threshold amplified spontaneous emission and lasing from colloidal
  nanocrystals of caesium lead halide perovskites},\ }\href
  {https://doi.org/10.1038/ncomms9056} {\bibfield  {journal} {\bibinfo
  {journal} {Nat.\ Commun.}\ }\textbf {\bibinfo {volume} {6}},\ \bibinfo
  {pages} {8056} (\bibinfo {year} {2015})}\BibitemShut {NoStop}%
\bibitem [{\citenamefont {Pan}\ \emph {et~al.}(2015)\citenamefont {Pan},
  \citenamefont {Sarmah}, \citenamefont {Murali}, \citenamefont {Dursun},
  \citenamefont {Peng}, \citenamefont {Parida}, \citenamefont {Liu},
  \citenamefont {Sinatra}, \citenamefont {Alyami}, \citenamefont {Zhao},
  \citenamefont {Alarousu}, \citenamefont {Ng}, \citenamefont {Ooi},
  \citenamefont {Bakr},\ and\ \citenamefont {Mohammed}}]{pan-15-psk}%
  \BibitemOpen
  \bibfield  {author} {\bibinfo {author} {\bibfnamefont {J.}~\bibnamefont
  {Pan}}, \bibinfo {author} {\bibfnamefont {S.~P.}\ \bibnamefont {Sarmah}},
  \bibinfo {author} {\bibfnamefont {B.}~\bibnamefont {Murali}}, \bibinfo
  {author} {\bibfnamefont {I.}~\bibnamefont {Dursun}}, \bibinfo {author}
  {\bibfnamefont {W.}~\bibnamefont {Peng}}, \bibinfo {author} {\bibfnamefont
  {M.~R.}\ \bibnamefont {Parida}}, \bibinfo {author} {\bibfnamefont
  {J.}~\bibnamefont {Liu}}, \bibinfo {author} {\bibfnamefont {L.}~\bibnamefont
  {Sinatra}}, \bibinfo {author} {\bibfnamefont {N.}~\bibnamefont {Alyami}},
  \bibinfo {author} {\bibfnamefont {C.}~\bibnamefont {Zhao}}, \bibinfo {author}
  {\bibfnamefont {E.}~\bibnamefont {Alarousu}}, \bibinfo {author}
  {\bibfnamefont {T.~K.}\ \bibnamefont {Ng}}, \bibinfo {author} {\bibfnamefont
  {B.~S.}\ \bibnamefont {Ooi}}, \bibinfo {author} {\bibfnamefont {O.~M.}\
  \bibnamefont {Bakr}},\ and\ \bibinfo {author} {\bibfnamefont {O.~F.}\
  \bibnamefont {Mohammed}},\ }\bibfield  {title} {\bibinfo {title} {Air-stable
  surface-passivated perovskite quantum dots for ultra-robust, single- and
  two-photon-induced amplified spontaneous emission},\ }\href
  {https://doi.org/10.1021/acs.jpclett.5b02460} {\bibfield  {journal} {\bibinfo
   {journal} {J.\ Phys.\ Chem.\ Lett.}\ }\textbf {\bibinfo {volume} {6}},\
  \bibinfo {pages} {5027} (\bibinfo {year} {2015})}\BibitemShut {NoStop}%
\bibitem [{\citenamefont {Utzat}\ \emph {et~al.}(2019)\citenamefont {Utzat},
  \citenamefont {Sun}, \citenamefont {Kaplan}, \citenamefont {Krieg},
  \citenamefont {Ginterseder}, \citenamefont {Spokoyny}, \citenamefont {Klein},
  \citenamefont {Shulenberger}, \citenamefont {Perkinson}, \citenamefont
  {Kovalenko},\ and\ \citenamefont {Bawendi}}]{utzat-19-psk}%
  \BibitemOpen
  \bibfield  {author} {\bibinfo {author} {\bibfnamefont {H.}~\bibnamefont
  {Utzat}}, \bibinfo {author} {\bibfnamefont {W.~W.}\ \bibnamefont {Sun}},
  \bibinfo {author} {\bibfnamefont {A.~E.~K.}\ \bibnamefont {Kaplan}}, \bibinfo
  {author} {\bibfnamefont {F.}~\bibnamefont {Krieg}}, \bibinfo {author}
  {\bibfnamefont {M.}~\bibnamefont {Ginterseder}}, \bibinfo {author}
  {\bibfnamefont {B.}~\bibnamefont {Spokoyny}}, \bibinfo {author}
  {\bibfnamefont {N.~D.}\ \bibnamefont {Klein}}, \bibinfo {author}
  {\bibfnamefont {K.~E.}\ \bibnamefont {Shulenberger}}, \bibinfo {author}
  {\bibfnamefont {C.~F.}\ \bibnamefont {Perkinson}}, \bibinfo {author}
  {\bibfnamefont {M.~V.}\ \bibnamefont {Kovalenko}},\ and\ \bibinfo {author}
  {\bibfnamefont {M.~G.}\ \bibnamefont {Bawendi}},\ }\bibfield  {title}
  {\bibinfo {title} {Coherent single-photon emission from colloidal lead halide
  perovskite quantum dots},\ }\href {https://doi.org/10.1126/science.aau7392}
  {\bibfield  {journal} {\bibinfo  {journal} {Science}\ }\textbf {\bibinfo
  {volume} {363}},\ \bibinfo {pages} {1068} (\bibinfo {year}
  {2019})}\BibitemShut {NoStop}%
\bibitem [{\citenamefont {Chen}\ \emph {et~al.}(2017)\citenamefont {Chen},
  \citenamefont {Zidek}, \citenamefont {Chabera}, \citenamefont {Liu},
  \citenamefont {Cheng}, \citenamefont {Nuuttila}, \citenamefont {Al-Marri},
  \citenamefont {Lehtivuori}, \citenamefont {Messing}, \citenamefont {Han},
  \citenamefont {Zheng},\ and\ \citenamefont {Pullerits}}]{chen-17-psk}%
  \BibitemOpen
  \bibfield  {author} {\bibinfo {author} {\bibfnamefont {J.~S.}\ \bibnamefont
  {Chen}}, \bibinfo {author} {\bibfnamefont {K.}~\bibnamefont {Zidek}},
  \bibinfo {author} {\bibfnamefont {P.}~\bibnamefont {Chabera}}, \bibinfo
  {author} {\bibfnamefont {D.~Z.}\ \bibnamefont {Liu}}, \bibinfo {author}
  {\bibfnamefont {P.~F.}\ \bibnamefont {Cheng}}, \bibinfo {author}
  {\bibfnamefont {L.}~\bibnamefont {Nuuttila}}, \bibinfo {author}
  {\bibfnamefont {M.~J.}\ \bibnamefont {Al-Marri}}, \bibinfo {author}
  {\bibfnamefont {H.}~\bibnamefont {Lehtivuori}}, \bibinfo {author}
  {\bibfnamefont {M.~E.}\ \bibnamefont {Messing}}, \bibinfo {author}
  {\bibfnamefont {K.~L.}\ \bibnamefont {Han}}, \bibinfo {author} {\bibfnamefont
  {K.~B.}\ \bibnamefont {Zheng}},\ and\ \bibinfo {author} {\bibfnamefont
  {T.}~\bibnamefont {Pullerits}},\ }\bibfield  {title} {\bibinfo {title} {Size-
  and wavelength-dependent two-photon absorption cross-section of {CsPbBr$_3$}
  perovskite quantum dots},\ }\href
  {https://doi.org/10.1021/acs.jpclett.7b00613} {\bibfield  {journal} {\bibinfo
   {journal} {J.\ Phys.\ Chem.\ Lett.}\ }\textbf {\bibinfo {volume} {8}},\
  \bibinfo {pages} {2316} (\bibinfo {year} {2017})}\BibitemShut {NoStop}%
\bibitem [{\citenamefont {Brennan}\ \emph {et~al.}(2017)\citenamefont
  {Brennan}, \citenamefont {Herr}, \citenamefont {Nguyen-Beck}, \citenamefont
  {Zinna}, \citenamefont {Draguta}, \citenamefont {Rouvimov}, \citenamefont
  {Parkhill},\ and\ \citenamefont {Kuno}}]{brennan-17-psk}%
  \BibitemOpen
  \bibfield  {author} {\bibinfo {author} {\bibfnamefont {M.~C.}\ \bibnamefont
  {Brennan}}, \bibinfo {author} {\bibfnamefont {J.~E.}\ \bibnamefont {Herr}},
  \bibinfo {author} {\bibfnamefont {T.~S.}\ \bibnamefont {Nguyen-Beck}},
  \bibinfo {author} {\bibfnamefont {J.}~\bibnamefont {Zinna}}, \bibinfo
  {author} {\bibfnamefont {S.}~\bibnamefont {Draguta}}, \bibinfo {author}
  {\bibfnamefont {S.}~\bibnamefont {Rouvimov}}, \bibinfo {author}
  {\bibfnamefont {J.}~\bibnamefont {Parkhill}},\ and\ \bibinfo {author}
  {\bibfnamefont {M.}~\bibnamefont {Kuno}},\ }\bibfield  {title} {\bibinfo
  {title} {Origin of the size-dependent {Stokes} shift in {CsPbBr$_3$}
  perovskite nanocrystals},\ }\href {https://doi.org/10.1021/jacs.7b05683}
  {\bibfield  {journal} {\bibinfo  {journal} {J.\ Am.\ Chem.\ Soc.}\ }\textbf
  {\bibinfo {volume} {139}},\ \bibinfo {pages} {12201} (\bibinfo {year}
  {2017})}\BibitemShut {NoStop}%
\bibitem [{\citenamefont {Efros}\ and\ \citenamefont
  {Efros}(1982)}]{efros-82-sqd}%
  \BibitemOpen
  \bibfield  {author} {\bibinfo {author} {\bibfnamefont {A.~L.}\ \bibnamefont
  {Efros}}\ and\ \bibinfo {author} {\bibfnamefont {A.~L.}\ \bibnamefont
  {Efros}},\ }\bibfield  {title} {\bibinfo {title} {Interband absorption of
  light in a semiconductor sphere},\ }\href@noop {} {\bibfield  {journal}
  {\bibinfo  {journal} {Sov.\ Phys.\ Semicond.}\ }\textbf {\bibinfo {volume}
  {16}},\ \bibinfo {pages} {772} (\bibinfo {year} {1982})}\BibitemShut
  {NoStop}%
\bibitem [{\citenamefont {Takagahara}(1987)}]{takagahara-87-sqd}%
  \BibitemOpen
  \bibfield  {author} {\bibinfo {author} {\bibfnamefont {T.}~\bibnamefont
  {Takagahara}},\ }\bibfield  {title} {\bibinfo {title} {Excitonic optical
  nonlinearity and exciton dynamics in semiconductor quantum dots},\ }\href
  {https://link.aps.org/doi/10.1103/PhysRevB.36.9293} {\bibfield  {journal}
  {\bibinfo  {journal} {Phys.\ Rev.\ B}\ }\textbf {\bibinfo {volume} {36}},\
  \bibinfo {pages} {9293} (\bibinfo {year} {1987})}\BibitemShut {NoStop}%
\bibitem [{\citenamefont {Nguyen}\ \emph
  {et~al.}(2020{\natexlab{a}})\citenamefont {Nguyen}, \citenamefont
  {Blundell},\ and\ \citenamefont {Guet}}]{nguyen-20b-psk}%
  \BibitemOpen
  \bibfield  {author} {\bibinfo {author} {\bibfnamefont {T.~P.~T.}\
  \bibnamefont {Nguyen}}, \bibinfo {author} {\bibfnamefont {S.~A.}\
  \bibnamefont {Blundell}},\ and\ \bibinfo {author} {\bibfnamefont
  {C.}~\bibnamefont {Guet}},\ }\bibfield  {title} {\bibinfo {title} {One-photon
  absorption by inorganic perovskite nanocrystals: A theoretical study},\
  }\href {https://doi.org/10.1103/PhysRevB.101.195414} {\bibfield  {journal}
  {\bibinfo  {journal} {Phys.\ Rev.\ B}\ }\textbf {\bibinfo {volume} {101}},\
  \bibinfo {pages} {195414} (\bibinfo {year} {2020}{\natexlab{a}})}\BibitemShut
  {NoStop}%
\bibitem [{\citenamefont {Chang}\ and\ \citenamefont
  {Xia}(1998)}]{chang-98-sqd}%
  \BibitemOpen
  \bibfield  {author} {\bibinfo {author} {\bibfnamefont {K.}~\bibnamefont
  {Chang}}\ and\ \bibinfo {author} {\bibfnamefont {J.~B.}\ \bibnamefont
  {Xia}},\ }\bibfield  {title} {\bibinfo {title} {Spatially separated excitons
  in quantum-dot quantum well structures},\ }\href
  {https://doi.org/10.1103/PhysRevB.57.9780} {\bibfield  {journal} {\bibinfo
  {journal} {Phys.\ Rev.\ B}\ }\textbf {\bibinfo {volume} {57}},\ \bibinfo
  {pages} {9780} (\bibinfo {year} {1998})}\BibitemShut {NoStop}%
\bibitem [{\citenamefont {Shumway}\ \emph {et~al.}(2001)\citenamefont
  {Shumway}, \citenamefont {Franceschetti},\ and\ \citenamefont
  {Zunger}}]{shumway-01-sqd}%
  \BibitemOpen
  \bibfield  {author} {\bibinfo {author} {\bibfnamefont {J.}~\bibnamefont
  {Shumway}}, \bibinfo {author} {\bibfnamefont {A.}~\bibnamefont
  {Franceschetti}},\ and\ \bibinfo {author} {\bibfnamefont {A.}~\bibnamefont
  {Zunger}},\ }\bibfield  {title} {\bibinfo {title} {Correlation versus
  mean-field contributions to excitons, multiexcitons, and charging energies in
  semiconductor quantum dots},\ }\href
  {https://doi.org/10.1103/PhysRevB.63.155316} {\bibfield  {journal} {\bibinfo
  {journal} {Phys.\ Rev.\ B}\ }\textbf {\bibinfo {volume} {63}},\ \bibinfo
  {pages} {155316} (\bibinfo {year} {2001})}\BibitemShut {NoStop}%
\bibitem [{\citenamefont {Tyrrell}\ and\ \citenamefont
  {Tomic}(2015)}]{tyrrell-15-sqd}%
  \BibitemOpen
  \bibfield  {author} {\bibinfo {author} {\bibfnamefont {E.~J.}\ \bibnamefont
  {Tyrrell}}\ and\ \bibinfo {author} {\bibfnamefont {S.}~\bibnamefont
  {Tomic}},\ }\bibfield  {title} {\bibinfo {title} {Effect of correlation and
  dielectric confinement on {$1S_{1/2}^{(e)} nS_{3/2}^{(h)}$} excitons in
  {CdTe/CdSe} and {CdSe/CdTe} type-{II} quantum dots},\ }\href
  {https://doi.org/10.1021/acs.jpcc.5b02789} {\bibfield  {journal} {\bibinfo
  {journal} {J.\ Phys.\ Chem.\ C}\ }\textbf {\bibinfo {volume} {119}},\
  \bibinfo {pages} {12720} (\bibinfo {year} {2015})}\BibitemShut {NoStop}%
\bibitem [{\citenamefont {Nguyen}\ \emph
  {et~al.}(2020{\natexlab{b}})\citenamefont {Nguyen}, \citenamefont
  {Blundell},\ and\ \citenamefont {Guet}}]{nguyen-20a-psk}%
  \BibitemOpen
  \bibfield  {author} {\bibinfo {author} {\bibfnamefont {T.~P.~T.}\
  \bibnamefont {Nguyen}}, \bibinfo {author} {\bibfnamefont {S.~A.}\
  \bibnamefont {Blundell}},\ and\ \bibinfo {author} {\bibfnamefont
  {C.}~\bibnamefont {Guet}},\ }\bibfield  {title} {\bibinfo {title}
  {Calculation of the biexciton shift in nanocrystals of inorganic
  perovskites},\ }\href {https://doi.org/10.1103/PhysRevB.101.125424}
  {\bibfield  {journal} {\bibinfo  {journal} {Phys.\ Rev.\ B}\ }\textbf
  {\bibinfo {volume} {101}},\ \bibinfo {pages} {125424} (\bibinfo {year}
  {2020}{\natexlab{b}})}\BibitemShut {NoStop}%
\bibitem [{\citenamefont {Sercel}\ \emph
  {et~al.}(2019{\natexlab{a}})\citenamefont {Sercel}, \citenamefont {Lyons},
  \citenamefont {Wickramaratne}, \citenamefont {Vaxenburg}, \citenamefont
  {Bernstein},\ and\ \citenamefont {Efros}}]{sercel-19-psk}%
  \BibitemOpen
  \bibfield  {author} {\bibinfo {author} {\bibfnamefont {P.~C.}\ \bibnamefont
  {Sercel}}, \bibinfo {author} {\bibfnamefont {J.~L.}\ \bibnamefont {Lyons}},
  \bibinfo {author} {\bibfnamefont {D.}~\bibnamefont {Wickramaratne}}, \bibinfo
  {author} {\bibfnamefont {R.}~\bibnamefont {Vaxenburg}}, \bibinfo {author}
  {\bibfnamefont {N.}~\bibnamefont {Bernstein}},\ and\ \bibinfo {author}
  {\bibfnamefont {A.~L.}\ \bibnamefont {Efros}},\ }\bibfield  {title} {\bibinfo
  {title} {Exciton fine structure in perovskite nanocrystals},\ }\href
  {https://doi.org/10.1021/acs.nanolett.9b01467} {\bibfield  {journal}
  {\bibinfo  {journal} {Nano Lett.}\ }\textbf {\bibinfo {volume} {19}},\
  \bibinfo {pages} {4068} (\bibinfo {year} {2019}{\natexlab{a}})}\BibitemShut
  {NoStop}%
\bibitem [{\citenamefont {Tamarat}\ \emph {et~al.}(2019)\citenamefont
  {Tamarat}, \citenamefont {Bodnarchuk}, \citenamefont {Trebbia}, \citenamefont
  {Erni}, \citenamefont {Kovalenko}, \citenamefont {Even},\ and\ \citenamefont
  {Lounis}}]{tamarat-19-psk}%
  \BibitemOpen
  \bibfield  {author} {\bibinfo {author} {\bibfnamefont {P.}~\bibnamefont
  {Tamarat}}, \bibinfo {author} {\bibfnamefont {M.~I.}\ \bibnamefont
  {Bodnarchuk}}, \bibinfo {author} {\bibfnamefont {J.-B.}\ \bibnamefont
  {Trebbia}}, \bibinfo {author} {\bibfnamefont {R.}~\bibnamefont {Erni}},
  \bibinfo {author} {\bibfnamefont {M.~V.}\ \bibnamefont {Kovalenko}}, \bibinfo
  {author} {\bibfnamefont {J.}~\bibnamefont {Even}},\ and\ \bibinfo {author}
  {\bibfnamefont {B.}~\bibnamefont {Lounis}},\ }\bibfield  {title} {\bibinfo
  {title} {The ground exciton state of formamidinium lead bromide perovskite
  nanocrystals is a singlet dark state},\ }\href
  {https://doi.org/10.1038/s41563-019-0364-x} {\bibfield  {journal} {\bibinfo
  {journal} {Nat.\ Mater.}\ }\textbf {\bibinfo {volume} {18}},\ \bibinfo
  {pages} {717} (\bibinfo {year} {2019})}\BibitemShut {NoStop}%
\bibitem [{\citenamefont {Kira}\ and\ \citenamefont {Koch}(2012)}]{Kira-Koch}%
  \BibitemOpen
  \bibfield  {author} {\bibinfo {author} {\bibfnamefont {M.}~\bibnamefont
  {Kira}}\ and\ \bibinfo {author} {\bibfnamefont {S.~W.}\ \bibnamefont
  {Koch}},\ }\href@noop {} {\emph {\bibinfo {title} {Semiconductor Quantum
  Optics}}}\ (\bibinfo  {publisher} {Cambridge University Press},\ \bibinfo
  {address} {New York},\ \bibinfo {year} {2012})\BibitemShut {NoStop}%
\bibitem [{\citenamefont {Efros}\ and\ \citenamefont
  {Rosen}(2000)}]{efros-00-sqd}%
  \BibitemOpen
  \bibfield  {author} {\bibinfo {author} {\bibfnamefont {A.~L.}\ \bibnamefont
  {Efros}}\ and\ \bibinfo {author} {\bibfnamefont {M.}~\bibnamefont {Rosen}},\
  }\bibfield  {title} {\bibinfo {title} {The electronic structure of
  semiconductor nanocrystals},\ }\href
  {https://doi.org/10.1146/annurev.matsci.30.1.475} {\bibfield  {journal}
  {\bibinfo  {journal} {Annu.\ Rev.\ Mater.\ Sci.}\ }\textbf {\bibinfo {volume}
  {30}},\ \bibinfo {pages} {475} (\bibinfo {year} {2000})}\BibitemShut
  {NoStop}%
\bibitem [{\citenamefont {Ekimov}\ \emph {et~al.}(1993)\citenamefont {Ekimov},
  \citenamefont {Hache}, \citenamefont {Schanneklein}, \citenamefont {Ricard},
  \citenamefont {Flytzanis}, \citenamefont {Kudryavtsev}, \citenamefont
  {Yazeva}, \citenamefont {Rodina},\ and\ \citenamefont
  {Efros}}]{ekimov-93-sqd}%
  \BibitemOpen
  \bibfield  {author} {\bibinfo {author} {\bibfnamefont {A.~I.}\ \bibnamefont
  {Ekimov}}, \bibinfo {author} {\bibfnamefont {F.}~\bibnamefont {Hache}},
  \bibinfo {author} {\bibfnamefont {M.~C.}\ \bibnamefont {Schanneklein}},
  \bibinfo {author} {\bibfnamefont {D.}~\bibnamefont {Ricard}}, \bibinfo
  {author} {\bibfnamefont {C.}~\bibnamefont {Flytzanis}}, \bibinfo {author}
  {\bibfnamefont {I.~A.}\ \bibnamefont {Kudryavtsev}}, \bibinfo {author}
  {\bibfnamefont {T.~V.}\ \bibnamefont {Yazeva}}, \bibinfo {author}
  {\bibfnamefont {A.~V.}\ \bibnamefont {Rodina}},\ and\ \bibinfo {author}
  {\bibfnamefont {A.~L.}\ \bibnamefont {Efros}},\ }\bibfield  {title} {\bibinfo
  {title} {Absorption and intensity-dependent photoluminescence measurements on
  {CdSe} quantum dots---assignment of the 1st electronic-transitions},\ }\href
  {http://dx.doi.org/10.1364/JOSAB.10.000100} {\bibfield  {journal} {\bibinfo
  {journal} {J.\ Opt.\ Soc.\ Am.\ B}\ }\textbf {\bibinfo {volume} {10}},\
  \bibinfo {pages} {100} (\bibinfo {year} {1993})}\BibitemShut {NoStop}%
\bibitem [{\citenamefont {Knox}(1963)}]{Knox}%
  \BibitemOpen
  \bibfield  {author} {\bibinfo {author} {\bibfnamefont {R.~S.}\ \bibnamefont
  {Knox}},\ }\href@noop {} {\emph {\bibinfo {title} {Theory of Excitons}}},\
  edited by\ \bibinfo {editor} {\bibfnamefont {F.}~\bibnamefont {Seitz}}\ and\
  \bibinfo {editor} {\bibfnamefont {D.}~\bibnamefont {Turnbull}},\ Solid State
  Physics, Supplement 5\ (\bibinfo  {publisher} {Academic},\ \bibinfo {address}
  {New York},\ \bibinfo {year} {1963})\BibitemShut {NoStop}%
\bibitem [{\citenamefont {Jackson}(1998)}]{Jackson}%
  \BibitemOpen
  \bibfield  {author} {\bibinfo {author} {\bibfnamefont {J.~D.}\ \bibnamefont
  {Jackson}},\ }\href@noop {} {\emph {\bibinfo {title} {Classical
  Electrodynamics}}},\ \bibinfo {edition} {3rd}\ ed.\ (\bibinfo  {publisher}
  {Wiley},\ \bibinfo {address} {New York},\ \bibinfo {year} {1998})\BibitemShut
  {NoStop}%
\bibitem [{\citenamefont {Karpulevich}\ \emph {et~al.}(2019)\citenamefont
  {Karpulevich}, \citenamefont {Bui}, \citenamefont {Wang}, \citenamefont
  {Hapke}, \citenamefont {Ramirez}, \citenamefont {Weller},\ and\ \citenamefont
  {Bester}}]{karpulevich-19-sqd}%
  \BibitemOpen
  \bibfield  {author} {\bibinfo {author} {\bibfnamefont {A.}~\bibnamefont
  {Karpulevich}}, \bibinfo {author} {\bibfnamefont {H.}~\bibnamefont {Bui}},
  \bibinfo {author} {\bibfnamefont {Z.}~\bibnamefont {Wang}}, \bibinfo {author}
  {\bibfnamefont {S.}~\bibnamefont {Hapke}}, \bibinfo {author} {\bibfnamefont
  {C.~P.}\ \bibnamefont {Ramirez}}, \bibinfo {author} {\bibfnamefont
  {H.}~\bibnamefont {Weller}},\ and\ \bibinfo {author} {\bibfnamefont
  {G.}~\bibnamefont {Bester}},\ }\bibfield  {title} {\bibinfo {title}
  {Dielectric response function for colloidal semiconductor quantum dots},\
  }\href {https://doi.org/10.1063/1.5128334} {\bibfield  {journal} {\bibinfo
  {journal} {J.\ Chem.\ Phys.}\ }\textbf {\bibinfo {volume} {151}},\ \bibinfo
  {pages} {224103} (\bibinfo {year} {2019})}\BibitemShut {NoStop}%
\bibitem [{\citenamefont {Pikus}\ and\ \citenamefont
  {Bir}(1971)}]{pikus-71-sqd}%
  \BibitemOpen
  \bibfield  {author} {\bibinfo {author} {\bibfnamefont {G.~E.}\ \bibnamefont
  {Pikus}}\ and\ \bibinfo {author} {\bibfnamefont {G.~L.}\ \bibnamefont
  {Bir}},\ }\bibfield  {title} {\bibinfo {title} {Exchange interaction in
  excitons in semiconductors},\ }\href
  {http://www.jetp.ac.ru/cgi-bin/dn/e_033_01_0108} {\bibfield  {journal}
  {\bibinfo  {journal} {Zh.\ Eksp.\ Teor.\ Fiz.}\ }\textbf {\bibinfo {volume}
  {60}},\ \bibinfo {pages} {195} (\bibinfo {year} {1971})}\BibitemShut
  {NoStop}%
\bibitem [{\citenamefont {Sercel}\ \emph
  {et~al.}(2019{\natexlab{b}})\citenamefont {Sercel}, \citenamefont {Lyons},
  \citenamefont {Bernstein},\ and\ \citenamefont {Efros}}]{sercel-19a-psk}%
  \BibitemOpen
  \bibfield  {author} {\bibinfo {author} {\bibfnamefont {P.~C.}\ \bibnamefont
  {Sercel}}, \bibinfo {author} {\bibfnamefont {J.~L.}\ \bibnamefont {Lyons}},
  \bibinfo {author} {\bibfnamefont {N.}~\bibnamefont {Bernstein}},\ and\
  \bibinfo {author} {\bibfnamefont {A.~L.}\ \bibnamefont {Efros}},\ }\bibfield
  {title} {\bibinfo {title} {Quasicubic model for metal halide perovskite
  nanocrystals},\ }\href {https://doi.org/10.1063/1.5127528} {\bibfield
  {journal} {\bibinfo  {journal} {J.\ Chem.\ Phys.}\ }\textbf {\bibinfo
  {volume} {151}},\ \bibinfo {pages} {234106} (\bibinfo {year}
  {2019}{\natexlab{b}})}\BibitemShut {NoStop}%
\bibitem [{\citenamefont {Blundell}\ \emph {et~al.}(2021)\citenamefont
  {Blundell}, \citenamefont {Nguyen},\ and\ \citenamefont
  {Guet}}]{blundell-21a-psk}%
  \BibitemOpen
  \bibfield  {author} {\bibinfo {author} {\bibfnamefont {S.~A.}\ \bibnamefont
  {Blundell}}, \bibinfo {author} {\bibfnamefont {T.~P.~T.}\ \bibnamefont
  {Nguyen}},\ and\ \bibinfo {author} {\bibfnamefont {C.}~\bibnamefont {Guet}},\
  }\bibfield  {title} {\bibinfo {title} {Calculation of two-photon absorption
  by nanocrystals of {${\mathrm{CsPbBr}}_{3}$}},\ }\href
  {https://link.aps.org/doi/10.1103/PhysRevB.103.045415} {\bibfield  {journal}
  {\bibinfo  {journal} {Phys.\ Rev.\ B}\ }\textbf {\bibinfo {volume} {103}},\
  \bibinfo {pages} {045415} (\bibinfo {year} {2021})}\BibitemShut {NoStop}%
\bibitem [{\citenamefont {Mahan}(2000)}]{Mahan}%
  \BibitemOpen
  \bibfield  {author} {\bibinfo {author} {\bibfnamefont {G.~D.}\ \bibnamefont
  {Mahan}},\ }\href@noop {} {\emph {\bibinfo {title} {Many-Particle
  Physics}}},\ \bibinfo {edition} {3rd}\ ed.\ (\bibinfo  {publisher} {Kluwer
  Academic/Plenum Publishers},\ \bibinfo {address} {New York},\ \bibinfo {year}
  {2000})\BibitemShut {NoStop}%
\bibitem [{\citenamefont {Amusia}\ and\ \citenamefont
  {Cherepkov}(1975)}]{amusia-75-sqd}%
  \BibitemOpen
  \bibfield  {author} {\bibinfo {author} {\bibfnamefont {M.~Y.}\ \bibnamefont
  {Amusia}}\ and\ \bibinfo {author} {\bibfnamefont {N.~A.}\ \bibnamefont
  {Cherepkov}},\ }\bibinfo {title} {Many-electron correlations in scattering
  processes},\ in\ \href@noop {} {\emph {\bibinfo {booktitle} {Case Studies in
  Atomic Physics}}},\ Vol.\ \bibinfo {volume} {5(2)}\ (\bibinfo  {publisher}
  {North-Holland},\ \bibinfo {address} {Amsterdam},\ \bibinfo {year} {1975})\
  pp.\ \bibinfo {pages} {47--179}\BibitemShut {NoStop}%
\bibitem [{\citenamefont {Guet}\ and\ \citenamefont
  {Johnson}(1992)}]{guet-92-sqd}%
  \BibitemOpen
  \bibfield  {author} {\bibinfo {author} {\bibfnamefont {C.}~\bibnamefont
  {Guet}}\ and\ \bibinfo {author} {\bibfnamefont {W.~R.}\ \bibnamefont
  {Johnson}},\ }\bibfield  {title} {\bibinfo {title} {Dipole excitations of
  closed-shell alkali-metal clusters},\ }\href
  {https://doi.org/10.1103/PhysRevB.45.11283} {\bibfield  {journal} {\bibinfo
  {journal} {Phys.\ Rev.\ B}\ }\textbf {\bibinfo {volume} {45}},\ \bibinfo
  {pages} {11283} (\bibinfo {year} {1992})}\BibitemShut {NoStop}%
\bibitem [{\citenamefont {Even}\ \emph {et~al.}(2014)\citenamefont {Even},
  \citenamefont {Pedesseau},\ and\ \citenamefont {Katan}}]{even-14b-psk}%
  \BibitemOpen
  \bibfield  {author} {\bibinfo {author} {\bibfnamefont {J.}~\bibnamefont
  {Even}}, \bibinfo {author} {\bibfnamefont {L.}~\bibnamefont {Pedesseau}},\
  and\ \bibinfo {author} {\bibfnamefont {C.}~\bibnamefont {Katan}},\ }\bibfield
   {title} {\bibinfo {title} {Analysis of multivalley and multibandgap
  absorption and enhancement of free carriers related to exciton screening in
  hybrid perovskites},\ }\href {http://dx.doi.org/10.1021/jp503337a} {\bibfield
   {journal} {\bibinfo  {journal} {J.\ Phys.\ Chem.\ C}\ }\textbf {\bibinfo
  {volume} {118}},\ \bibinfo {pages} {11566} (\bibinfo {year}
  {2014})}\BibitemShut {NoStop}%
\bibitem [{\citenamefont {Yang}\ \emph {et~al.}(2017)\citenamefont {Yang},
  \citenamefont {Surrente}, \citenamefont {Galkowski}, \citenamefont {Miyata},
  \citenamefont {Portugall}, \citenamefont {Sutton}, \citenamefont
  {Haghighirad}, \citenamefont {Snaith}, \citenamefont {Maude}, \citenamefont
  {Plochocka},\ and\ \citenamefont {Nicholas}}]{yang-17-psk}%
  \BibitemOpen
  \bibfield  {author} {\bibinfo {author} {\bibfnamefont {Z.}~\bibnamefont
  {Yang}}, \bibinfo {author} {\bibfnamefont {A.}~\bibnamefont {Surrente}},
  \bibinfo {author} {\bibfnamefont {K.}~\bibnamefont {Galkowski}}, \bibinfo
  {author} {\bibfnamefont {A.}~\bibnamefont {Miyata}}, \bibinfo {author}
  {\bibfnamefont {O.}~\bibnamefont {Portugall}}, \bibinfo {author}
  {\bibfnamefont {R.~J.}\ \bibnamefont {Sutton}}, \bibinfo {author}
  {\bibfnamefont {A.~A.}\ \bibnamefont {Haghighirad}}, \bibinfo {author}
  {\bibfnamefont {H.~J.}\ \bibnamefont {Snaith}}, \bibinfo {author}
  {\bibfnamefont {D.~K.}\ \bibnamefont {Maude}}, \bibinfo {author}
  {\bibfnamefont {P.}~\bibnamefont {Plochocka}},\ and\ \bibinfo {author}
  {\bibfnamefont {R.~J.}\ \bibnamefont {Nicholas}},\ }\bibfield  {title}
  {\bibinfo {title} {Impact of the halide cage on the electronic properties of
  fully inorganic cesium lead halide perovskites},\ }\href
  {https://doi.org/10.1021/acsenergylett.7b00416} {\bibfield  {journal}
  {\bibinfo  {journal} {ACS Energy Lett.}\ }\textbf {\bibinfo {volume} {2}},\
  \bibinfo {pages} {1621} (\bibinfo {year} {2017})}\BibitemShut {NoStop}%
\bibitem [{\citenamefont {Fu}\ \emph {et~al.}(2017{\natexlab{a}})\citenamefont
  {Fu}, \citenamefont {Xu}, \citenamefont {Han}, \citenamefont {Wu},
  \citenamefont {Huan}, \citenamefont {Leek},\ and\ \citenamefont
  {Sum}}]{fu-17a-psk}%
  \BibitemOpen
  \bibfield  {author} {\bibinfo {author} {\bibfnamefont {J.}~\bibnamefont
  {Fu}}, \bibinfo {author} {\bibfnamefont {Q.}~\bibnamefont {Xu}}, \bibinfo
  {author} {\bibfnamefont {G.}~\bibnamefont {Han}}, \bibinfo {author}
  {\bibfnamefont {B.}~\bibnamefont {Wu}}, \bibinfo {author} {\bibfnamefont
  {C.~H.~A.}\ \bibnamefont {Huan}}, \bibinfo {author} {\bibfnamefont {M.~L.}\
  \bibnamefont {Leek}},\ and\ \bibinfo {author} {\bibfnamefont {T.~C.}\
  \bibnamefont {Sum}},\ }\bibfield  {title} {\bibinfo {title} {Hot carrier
  cooling mechanisms in halide perovskites},\ }\href
  {https://www.nature.com/articles/s41467-017-01360-3} {\bibfield  {journal}
  {\bibinfo  {journal} {Nat.\ Commun.}\ }\textbf {\bibinfo {volume} {8}},\
  \bibinfo {pages} {1} (\bibinfo {year} {2017}{\natexlab{a}})}\BibitemShut
  {NoStop}%
\bibitem [{\citenamefont {Shcherbakov-Wu}\ \emph {et~al.}(2021)\citenamefont
  {Shcherbakov-Wu}, \citenamefont {Sercel}, \citenamefont {Krieg},
  \citenamefont {Kovalenko},\ and\ \citenamefont
  {Tisdale}}]{shcherbakov-wu-21-psk}%
  \BibitemOpen
  \bibfield  {author} {\bibinfo {author} {\bibfnamefont {W.}~\bibnamefont
  {Shcherbakov-Wu}}, \bibinfo {author} {\bibfnamefont {P.~C.}\ \bibnamefont
  {Sercel}}, \bibinfo {author} {\bibfnamefont {F.}~\bibnamefont {Krieg}},
  \bibinfo {author} {\bibfnamefont {M.~V.}\ \bibnamefont {Kovalenko}},\ and\
  \bibinfo {author} {\bibfnamefont {W.~A.}\ \bibnamefont {Tisdale}},\
  }\bibfield  {title} {\bibinfo {title} {Temperature-independent dielectric
  constant in {CsPbBr$_3$} nanocrystals revealed by linear absorption
  spectroscopy},\ }\href {https://doi.org/10.1021/acs.jpclett.1c01822}
  {\bibfield  {journal} {\bibinfo  {journal} {J.\ Phys.\ Chem.\ Lett.}\
  }\textbf {\bibinfo {volume} {12}},\ \bibinfo {pages} {8088} (\bibinfo {year}
  {2021})}\BibitemShut {NoStop}%
\bibitem [{\citenamefont {Dirin}\ \emph {et~al.}(2016)\citenamefont {Dirin},
  \citenamefont {Cherniukh}, \citenamefont {Yakunin}, \citenamefont
  {Shynkarenko},\ and\ \citenamefont {Kovalenko}}]{dirin-16-psk}%
  \BibitemOpen
  \bibfield  {author} {\bibinfo {author} {\bibfnamefont {D.~N.}\ \bibnamefont
  {Dirin}}, \bibinfo {author} {\bibfnamefont {I.}~\bibnamefont {Cherniukh}},
  \bibinfo {author} {\bibfnamefont {S.}~\bibnamefont {Yakunin}}, \bibinfo
  {author} {\bibfnamefont {Y.}~\bibnamefont {Shynkarenko}},\ and\ \bibinfo
  {author} {\bibfnamefont {M.~V.}\ \bibnamefont {Kovalenko}},\ }\bibfield
  {title} {\bibinfo {title} {Solution-grown {CsPbBr$_{3}$} perovskite single
  crystals for photon detection},\ }\href
  {https://doi.org/10.1021/acs.chemmater.6b04298} {\bibfield  {journal}
  {\bibinfo  {journal} {Chem.\ Mater.}\ }\textbf {\bibinfo {volume} {28}},\
  \bibinfo {pages} {8470} (\bibinfo {year} {2016})}\BibitemShut {NoStop}%
\bibitem [{\citenamefont {Nguyen}(2020)}]{nguyen-20c-psk}%
  \BibitemOpen
  \bibfield  {author} {\bibinfo {author} {\bibfnamefont {T.~P.~T.}\
  \bibnamefont {Nguyen}},\ }\emph {\bibinfo {title} {A theoretical study of
  correlation effects of {N} electrons in semiconductor nanocrystals:
  Applications to optoelectronic properties of perovskite nanocrystals}},\
  \href@noop {} {Ph.D. thesis},\ \bibinfo  {school} {Univ.\ Grenoble Alpes},
  \bibinfo {address} {France} (\bibinfo {year} {2020})\BibitemShut {NoStop}%
\bibitem [{\citenamefont {Fu}\ \emph {et~al.}(2017{\natexlab{b}})\citenamefont
  {Fu}, \citenamefont {Tamarat}, \citenamefont {Huang}, \citenamefont {Even},
  \citenamefont {Rogach},\ and\ \citenamefont {Lounis}}]{fu-17-psk}%
  \BibitemOpen
  \bibfield  {author} {\bibinfo {author} {\bibfnamefont {M.}~\bibnamefont
  {Fu}}, \bibinfo {author} {\bibfnamefont {P.}~\bibnamefont {Tamarat}},
  \bibinfo {author} {\bibfnamefont {H.}~\bibnamefont {Huang}}, \bibinfo
  {author} {\bibfnamefont {J.}~\bibnamefont {Even}}, \bibinfo {author}
  {\bibfnamefont {A.~L.}\ \bibnamefont {Rogach}},\ and\ \bibinfo {author}
  {\bibfnamefont {B.}~\bibnamefont {Lounis}},\ }\bibfield  {title} {\bibinfo
  {title} {Neutral and charged exciton fine structure in single lead halide
  perovskite nanocrystals revealed by magneto-optical spectroscopy},\ }\href
  {https://doi.org/10.1021/acs.nanolett.7b00064} {\bibfield  {journal}
  {\bibinfo  {journal} {Nano Lett.}\ }\textbf {\bibinfo {volume} {17}},\
  \bibinfo {pages} {2895} (\bibinfo {year} {2017}{\natexlab{b}})}\BibitemShut
  {NoStop}%
\bibitem [{\citenamefont {Canneson}\ \emph {et~al.}(2017)\citenamefont
  {Canneson}, \citenamefont {Shornikova}, \citenamefont {Yakovlev},
  \citenamefont {Rogge}, \citenamefont {Mitioglu}, \citenamefont {Ballottin},
  \citenamefont {Christianen}, \citenamefont {Lhuillier}, \citenamefont
  {Bayer},\ and\ \citenamefont {Biadala}}]{canneson-17-psk}%
  \BibitemOpen
  \bibfield  {author} {\bibinfo {author} {\bibfnamefont {D.}~\bibnamefont
  {Canneson}}, \bibinfo {author} {\bibfnamefont {E.~V.}\ \bibnamefont
  {Shornikova}}, \bibinfo {author} {\bibfnamefont {D.~R.}\ \bibnamefont
  {Yakovlev}}, \bibinfo {author} {\bibfnamefont {T.}~\bibnamefont {Rogge}},
  \bibinfo {author} {\bibfnamefont {A.~A.}\ \bibnamefont {Mitioglu}}, \bibinfo
  {author} {\bibfnamefont {M.~V.}\ \bibnamefont {Ballottin}}, \bibinfo {author}
  {\bibfnamefont {P.~C.~M.}\ \bibnamefont {Christianen}}, \bibinfo {author}
  {\bibfnamefont {E.}~\bibnamefont {Lhuillier}}, \bibinfo {author}
  {\bibfnamefont {M.}~\bibnamefont {Bayer}},\ and\ \bibinfo {author}
  {\bibfnamefont {L.}~\bibnamefont {Biadala}},\ }\bibfield  {title} {\bibinfo
  {title} {Negatively charged and dark excitons in {CsPbBr$_{3}$} perovskite
  nanocrystals revealed by high magnetic fields},\ }\href
  {https://doi.org/10.1021/acs.nanolett.7b02827} {\bibfield  {journal}
  {\bibinfo  {journal} {Nano Lett.}\ }\textbf {\bibinfo {volume} {17}},\
  \bibinfo {pages} {6177} (\bibinfo {year} {2017})}\BibitemShut {NoStop}%
\bibitem [{\citenamefont {Ben~Aich}\ \emph {et~al.}(2019)\citenamefont
  {Ben~Aich}, \citenamefont {Saidi}, \citenamefont {Ben~Radhia}, \citenamefont
  {Boujdaria}, \citenamefont {Barisien}, \citenamefont {Legrand}, \citenamefont
  {Bernardot}, \citenamefont {Chamarro},\ and\ \citenamefont
  {Testelin}}]{ben-aich-19-psk}%
  \BibitemOpen
  \bibfield  {author} {\bibinfo {author} {\bibfnamefont {R.}~\bibnamefont
  {Ben~Aich}}, \bibinfo {author} {\bibfnamefont {I.}~\bibnamefont {Saidi}},
  \bibinfo {author} {\bibfnamefont {S.}~\bibnamefont {Ben~Radhia}}, \bibinfo
  {author} {\bibfnamefont {K.}~\bibnamefont {Boujdaria}}, \bibinfo {author}
  {\bibfnamefont {T.}~\bibnamefont {Barisien}}, \bibinfo {author}
  {\bibfnamefont {L.}~\bibnamefont {Legrand}}, \bibinfo {author} {\bibfnamefont
  {F.}~\bibnamefont {Bernardot}}, \bibinfo {author} {\bibfnamefont
  {M.}~\bibnamefont {Chamarro}},\ and\ \bibinfo {author} {\bibfnamefont
  {C.}~\bibnamefont {Testelin}},\ }\bibfield  {title} {\bibinfo {title}
  {Bright-exciton splittings in inorganic cesium lead halide perovskite
  nanocrystals},\ }\href {https://doi.org/10.1103/PhysRevApplied.11.034042}
  {\bibfield  {journal} {\bibinfo  {journal} {Phys.\ Rev.\ Appl.}\ }\textbf
  {\bibinfo {volume} {11}},\ \bibinfo {pages} {034042} (\bibinfo {year}
  {2019})}\BibitemShut {NoStop}%
\bibitem [{\citenamefont {Ben~Aich}\ \emph {et~al.}(2020)\citenamefont
  {Ben~Aich}, \citenamefont {Ben~Radhia}, \citenamefont {Boujdaria},
  \citenamefont {Chamarro},\ and\ \citenamefont {Testelin}}]{ben-aich-20-psk}%
  \BibitemOpen
  \bibfield  {author} {\bibinfo {author} {\bibfnamefont {R.}~\bibnamefont
  {Ben~Aich}}, \bibinfo {author} {\bibfnamefont {S.}~\bibnamefont
  {Ben~Radhia}}, \bibinfo {author} {\bibfnamefont {K.}~\bibnamefont
  {Boujdaria}}, \bibinfo {author} {\bibfnamefont {M.}~\bibnamefont
  {Chamarro}},\ and\ \bibinfo {author} {\bibfnamefont {C.}~\bibnamefont
  {Testelin}},\ }\bibfield  {title} {\bibinfo {title} {Multiband {$\mathbf{k}
  \cdot \mathbf{p}$} model for tetragonal crystals: Application to hybrid
  halide perovskite nanocrystals},\ }\href
  {https://doi.org/10.1021/acs.jpclett.9b02179} {\bibfield  {journal} {\bibinfo
   {journal} {J.\ Phys.\ Chem.\ Lett.}\ }\textbf {\bibinfo {volume} {11}},\
  \bibinfo {pages} {808} (\bibinfo {year} {2020})}\BibitemShut {NoStop}%
\bibitem [{\citenamefont {Swift}\ \emph {et~al.}(2021)\citenamefont {Swift},
  \citenamefont {Lyons}, \citenamefont {Efros},\ and\ \citenamefont
  {Sercel}}]{swift-21-psk}%
  \BibitemOpen
  \bibfield  {author} {\bibinfo {author} {\bibfnamefont {M.~W.}\ \bibnamefont
  {Swift}}, \bibinfo {author} {\bibfnamefont {J.~L.}\ \bibnamefont {Lyons}},
  \bibinfo {author} {\bibfnamefont {A.~L.}\ \bibnamefont {Efros}},\ and\
  \bibinfo {author} {\bibfnamefont {P.~C.}\ \bibnamefont {Sercel}},\ }\bibfield
   {title} {\bibinfo {title} {Rashba exciton in a {2D} perovskite quantum
  dot},\ }\href {https://doi.org/10.1039/d1nr04884h} {\bibfield  {journal}
  {\bibinfo  {journal} {Nanoscale}\ }\textbf {\bibinfo {volume} {13}},\
  \bibinfo {pages} {16769} (\bibinfo {year} {2021})}\BibitemShut {NoStop}%
\bibitem [{\citenamefont {Takagahara}(1993)}]{takagahara-93-sqd}%
  \BibitemOpen
  \bibfield  {author} {\bibinfo {author} {\bibfnamefont {T.}~\bibnamefont
  {Takagahara}},\ }\bibfield  {title} {\bibinfo {title} {Effects of dielectric
  confinement and electron-hole exchange interaction on excitonic states in
  semiconductor quantum dots},\ }\href
  {https://link.aps.org/doi/10.1103/PhysRevB.47.4569} {\bibfield  {journal}
  {\bibinfo  {journal} {Phys.\ Rev.\ B}\ }\textbf {\bibinfo {volume} {47}},\
  \bibinfo {pages} {4569} (\bibinfo {year} {1993})}\BibitemShut {NoStop}%
\bibitem [{\citenamefont {Tong}\ and\ \citenamefont {Wu}(2011)}]{tong-11-sqd}%
  \BibitemOpen
  \bibfield  {author} {\bibinfo {author} {\bibfnamefont {H.}~\bibnamefont
  {Tong}}\ and\ \bibinfo {author} {\bibfnamefont {M.~W.}\ \bibnamefont {Wu}},\
  }\bibfield  {title} {\bibinfo {title} {Theory of excitons in cubic {III-V}
  semiconductor {GaAs, InAs and GaN} quantum dots: Fine structure and spin
  relaxation},\ }\href {https://doi.org/10.1103/PhysRevB.83.235323} {\bibfield
  {journal} {\bibinfo  {journal} {Phys.\ Rev.\ B}\ }\textbf {\bibinfo {volume}
  {83}},\ \bibinfo {pages} {235323} (\bibinfo {year} {2011})}\BibitemShut
  {NoStop}%
\bibitem [{\citenamefont {Elliott}(1957)}]{elliott-57-sqd}%
  \BibitemOpen
  \bibfield  {author} {\bibinfo {author} {\bibfnamefont {R.~J.}\ \bibnamefont
  {Elliott}},\ }\bibfield  {title} {\bibinfo {title} {Intensity of optical
  absorption by excitons},\ }\href {https://doi.org/10.1103/PhysRev.108.1384}
  {\bibfield  {journal} {\bibinfo  {journal} {Phys.\ Rev.}\ }\textbf {\bibinfo
  {volume} {108}},\ \bibinfo {pages} {1384} (\bibinfo {year}
  {1957})}\BibitemShut {NoStop}%
\bibitem [{\citenamefont {Tanaka}\ \emph {et~al.}(2003)\citenamefont {Tanaka},
  \citenamefont {Takahashi}, \citenamefont {Ban}, \citenamefont {Kondo},
  \citenamefont {Uchida},\ and\ \citenamefont {Miura}}]{tanaka-03-psk}%
  \BibitemOpen
  \bibfield  {author} {\bibinfo {author} {\bibfnamefont {K.}~\bibnamefont
  {Tanaka}}, \bibinfo {author} {\bibfnamefont {T.}~\bibnamefont {Takahashi}},
  \bibinfo {author} {\bibfnamefont {T.}~\bibnamefont {Ban}}, \bibinfo {author}
  {\bibfnamefont {T.}~\bibnamefont {Kondo}}, \bibinfo {author} {\bibfnamefont
  {K.}~\bibnamefont {Uchida}},\ and\ \bibinfo {author} {\bibfnamefont
  {N.}~\bibnamefont {Miura}},\ }\bibfield  {title} {\bibinfo {title}
  {Comparative study on the excitons in lead-halide-based perovskite-type
  crystals {CH$_3$NH$_3$PbBr$_3$} {CH$_3$NH$_3$PbI$_3$}},\ }\href
  {https://doi.org/10.1016/S0038-1098(03)00566-0} {\bibfield  {journal}
  {\bibinfo  {journal} {Solid State Commun.}\ }\textbf {\bibinfo {volume}
  {127}},\ \bibinfo {pages} {619} (\bibinfo {year} {2003})}\BibitemShut
  {NoStop}%
\bibitem [{\citenamefont {Brink}\ and\ \citenamefont
  {Satchler}(1994)}]{Brink-Satchler}%
  \BibitemOpen
  \bibfield  {author} {\bibinfo {author} {\bibfnamefont {D.~M.}\ \bibnamefont
  {Brink}}\ and\ \bibinfo {author} {\bibfnamefont {G.~R.}\ \bibnamefont
  {Satchler}},\ }\href@noop {} {\emph {\bibinfo {title} {Angular Momentum}}},\
  \bibinfo {edition} {3rd}\ ed.\ (\bibinfo  {publisher} {Clarendon Press},\
  \bibinfo {address} {Oxford},\ \bibinfo {year} {1994})\BibitemShut {NoStop}%
\end{thebibliography}
\end{document}